# High-Order harmonics of Thermal Tides observed in the atmosphere of Mars by the Pressure Sensor on the Insight lander

J. Hernández-Bernal[1], A. Spiga[1], F. Forget[1], D. Banfield[2]

[1]Laboratoire de Météorologie Dynamique, Sorbonne Université, Paris, France

[2]NASA Ames Research Center

Corresponding author: Jorge Hernández Bernal

jorge.hernandez-bernal@lmd.ipsl.fr

**Key Points:**

- Analysis of an unprecedented dataset of pressure obtained by Insight suggests that tidal harmonics beyond 24 are present on Mars.
- Even and odd modes exhibit distinct patterns with a seasonal dependency centered on equinoxes and solstices, and response to dust events.

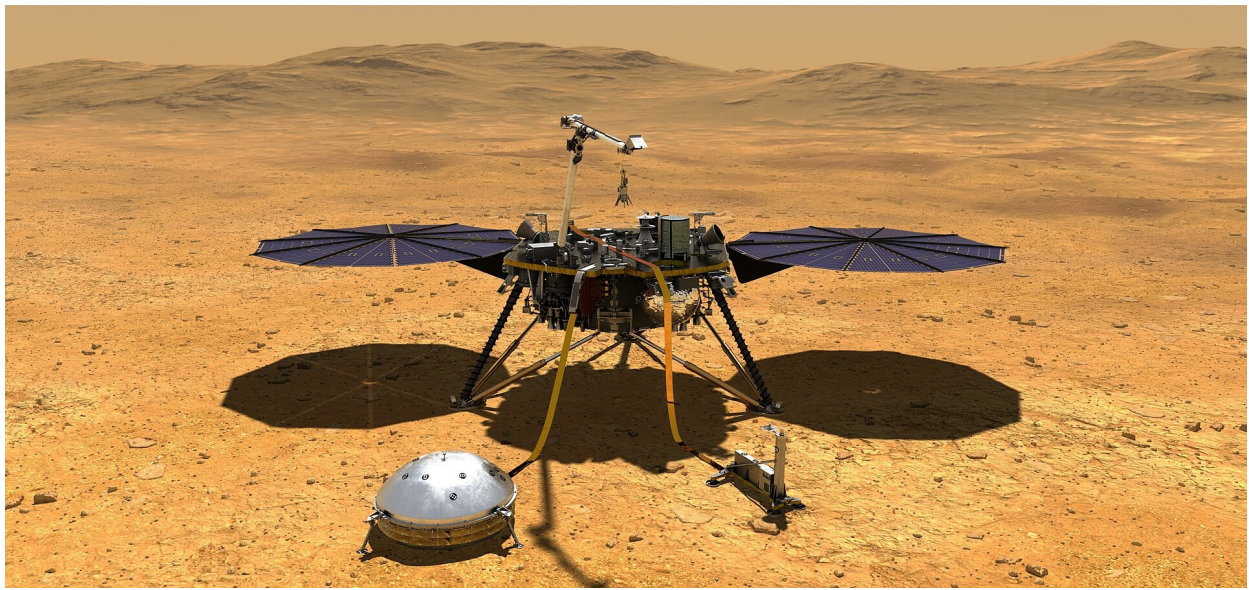

**Abstract**

Thermal tides are atmospheric planetary-scale waves with periods that are harmonics of the solar day. In the Martian atmosphere thermal tides are known to be especially significant compared to any other known planet. Based on the dataset of pressure timeseries produced by the Insight lander, which is unprecedented in terms of accuracy and temporal coverage, we investigate thermal tides on Mars and we find harmonics even beyond the number 24, which exceeds significantly the number of harmonics previously reported by other works. We explore comparatively the characteristics and seasonal evolution of tidal harmonics and find that even and odd harmonics exhibit some clearly differentiated trends that evolve seasonally and respond to dust events. High-order tidal harmonics with small amplitudes could transiently interfere constructively to produce meteorologically relevant patterns.

**Plain Language Summary**

In analogy to the string of a guitar, which can oscillate in integer harmonics, planetary atmospheres exhibit oscillations that are harmonics of the solar day (Harmonic 1 with a period of 24h; harmonic 2, 12h; harmonic 3, 8h; etc.). These oscillations are part of the so-called "atmospheric thermal tides", which retain a complex global structure. They are conceptually related to ocean gravitational tides, and they have been observed in atmospheres of the solar system whose main source of energy is the light from the sun: Earth, Mars, Venus, and Titan. On Mars, thermal tides are particularly strong and they play a key role in atmospheric dynamics, presenting interactions with meteorological phenomena such as dust storms. Most studies on thermal tides focus on low-order harmonics (1,2,3, and sometimes 4). In this study we use a particularly sensitive pressure sensor that landed on Mars with the Insight mission to explore the existence of high-order harmonics, and we find clear harmonics with very small amplitudes even beyond harmonic 24, which corresponds to 24 oscillations per solar day. We compare the characteristics of those harmonics and analyze their seasonal behavior, and we find that even and odd harmonics exhibit clearly different behaviors.

## 1 Introduction

Atmospheric tides are a natural response of planetary atmospheres to the periodic forcing exerted whether by gravity (gravitational tides), or insolation (thermal tides). They consist of planetary-scale oscillation modes with periods that are harmonics of the forcing period (the solar rotation period, in the case of thermal tides) and integer zonal wavenumbers. A tide with an harmonic number equal to the zonal wavenumber has a phase speed equal to the apparent speed of the Sun, and is called a migrating tide. Other tides with various phase speeds are called non-migrating tides. A detailed explanation of migrating and non-migrating tides can be found for example in Forbes et al. (2020). The classical theory of atmospheric tides can be found in Chapman & Lindzen (1969).

Thermal tides have been observed in the atmospheres of the solar system whose primary source of energy is insolation: Earth (Chapman & Lidzen, 1969), Mars (e.g. Hess et al., 1977; Forbes et al., 2020), Venus (Peralta et al., 2012; Kouyama et al., 2019), and Titan (Tokano, 2010). They are part of the general circulation, interact with atmospheric phenomena, and link the neutral lower and ionized upper atmosphere (Schindelegger et al., 2023).

Most studies focus on tidal harmonics $\mathcal{S}1$ and $\mathcal{S}2$ (i.e. harmonics 1 and 2, we will use this notation from now onwards), given that low order harmonics are much stronger compared to higher ones. Some detailed studies have also analyzed $\mathcal{S}3$ and $\mathcal{S}4$ on Earth (e.g. Moudden & Forbes, 2013; Smith et al., 2014) and reported them in Venus and in Mars (Peralta et al., 2012; Guzewich et al., 2016). Higher order harmonics have been detected on Earth (up to $\mathcal{S}12$; Sakazaki and Hamilton, 2020; Hupe et al., 2018; Hedlin et al., 2018; He et al., 2020) and on Mars (up to $\mathcal{S}6$; Sánchez-Lavega et al., 2022). In addition to it, recent studies based on simulations (Lian et al., 2023; Wilson et al., 2017; Wilson et al., 2015) have found harmonics up to $\mathcal{S}6$ and $\mathcal{S}7$ on Mars, and they indicate that such harmonics are dominated by migrating modes. On Earth, other mechanisms in addition to solar insolation have been observed to contribute to $\mathcal{S}3$ and $\mathcal{S}4$: nonlinear interaction between tides (e.g. Teitelbaum et al., 1989), and interaction between tides and gravity waves (e.g. Geißler et al., 2020), other references can be found in Pancheva et al. (2021).

On Mars, due to the low thermal inertia of the atmosphere, thermal tides are stronger in relation to the atmospheric thickness than in any other planet of the solar system (Barnes et al, 2017); they are a key aspect of the general circulation, and they react strongly to the presence of dust (Leovy and Zurek, 1979; Ordóñez-Etxeberria et al., 2019, Viúdez-Moreiras et al., 2020), leading to feedback that could play a key role in the establishment of large dust events (Barnes et al., 2017). Many authors have explored thermal tides on Mars based on surface stations (Hess et al., 1977; Guzewich et al., 2016; Sánchez-Lavega et al., 2022; Viúdez-Moreiras et al., 2020; this work), satellites (e.g. Whiters et al, 2011; Forbes et al., 2020; López-Valverde et al., 2023; Fan et al., 2022; Guerlet et al., 2023), and simulations (e.g. Wilson and Hamilton, 1996; Guerlet et al., 2023). Due to their lower sensibility compared to ground stations, remote sensing instruments from satellites can only detect the stronger low-order tidal harmonics. As a counterpart, a single station on the ground can observe the daily cycle at a single location, which is insufficient to separate modes with different wavenumbers. Therefore, a given harmonic number observed by a ground station is actually a mix of tidal modes with different zonal wavenumbers. This is why a network of ground stations on Mars would be especially useful to disentangle the global structure of thermal tides (e.g. Wilson and Kahre, 2022).

In this study, we benefit from the unprecedented accuracy (50mPa) and temporal coverage of the dataset produced by the Pressure Sensor (PS) on the Insight lander (Spiga et al., 2018; Banfield et al., 2019; Banfield, Spiga, et al. 2020) to investigate thermal tides on Mars from Ls (Solar Longitude) 304º in MY34 (Martian Year 34), to Ls 20º in MY36. This interval corresponds to sols 15-824 of the Insight mission. Our results support the existence of tidal modes beyond $\mathcal{S}24$. We comparatively analyze the characteristics and seasonal evolution of such harmonics. Results are discussed in relation to dust climatologies produced by Montabone et al. (2015; 2020), and using the nomenclature for annually repeating dust events (A, B, C) proposed by Kass et al. (2016). The main dust events that took place during the acquisition of this dataset are: dust event C in MY34 (sols 40-80; C34 from now onwards; analyzed in detail by Viúdez-Moreiras et al., 2020), regional off-season dust event in MY35 (sols 180-220; R), dust event A in MY35 (sols 550-650; A35), and dust event C in MY35 (sols 700-740; C35).

For convenience, we use a nomenclature for seasons centered on equinoxes and solstices (with a shift of 45º in Solar Longitude (Ls) from the conventional meteorological seasons). In the context of this paper, the northward equinox season refers to Ls 315º-45º, northern solstice season refers to Ls 45º-135º, southward equinox season refers to Ls 135º-215º, and southern solstice season refers to Ls 225º-315.

## 2 Results

In this section we discuss the number of tidal harmonics present in the dataset (subsection 2.1), show the overall differences between even and odd harmonics (subsection 2.2), explore the seasonal evolution of the diurnal pressure cycle driven by the ensemble of tidal harmonics (subsection 2.3), and look in detail at the properties of separate harmonics (subsection 2.4). We use Numpy (Harris et al., 2020) and Scipy (Virtanen et al., 2020) for our computations.

## 2.1 High-order tidal harmonics in the Insight PS dataset

We compute a periodogram (as implemented by Virtanen et al., 2020) on the longest interval with continuous Insight measurements available. We show this periodogram in fig. 1.

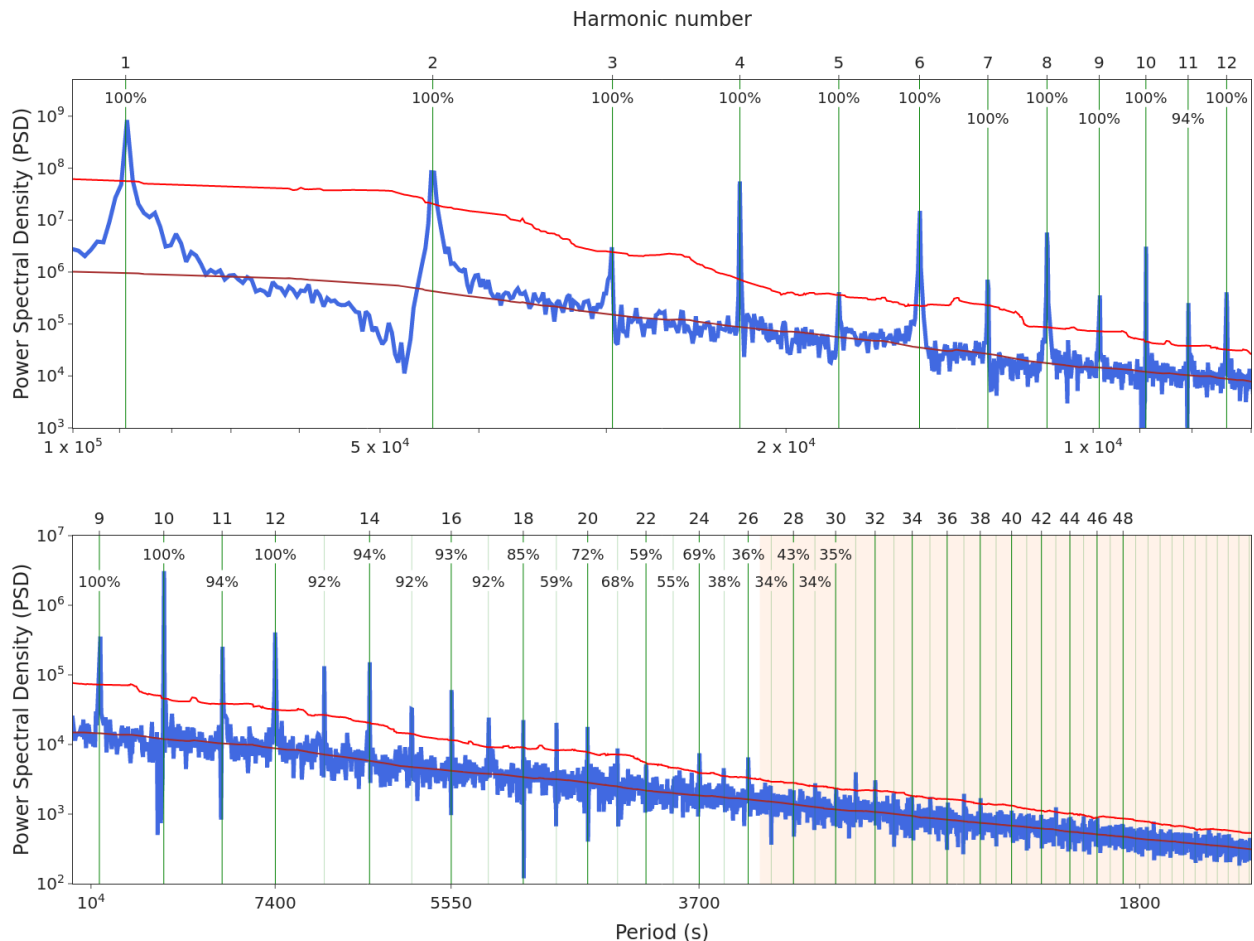

Figure 1. Periodogram showing Power Spectral Density (PSD) as function of wave period computed on an interval with continuous coverage over 77 sols (sols 490-567). Bottom graph is simply a continuation of the top one. Harmonics of the Mars mean solar rotation period (~88775s) are indicated by green vertical lines with the harmonic number written in the upper axis. The percentage of sols within this interval in which the amplitude of each harmonic is over instrumental noise level is indicated in the top of the graphs. Brown curve represents the average PSD, red line represents an indicative estimation of the noise level (see supporting text S2 for details). Yellow shadowing starting after harmonic number 26 represents a part of the periodogram where some PSD peaks coincident with harmonics are coincident but it is not clear if they are actual tidal harmonics. The correction proposed by Lange et al. (2022) for the PS dataset does not produce any significant change in this periodogram.

Clear PSD peaks over the estimated noise level coincident with expected tidal harmonics are present up to $\mathcal{S}26$, being the absence of a peak for $\mathcal{S}23$ the only exception. Other tidal harmonics might be present beyond $\mathcal{S}26$, clear peaks are $\mathcal{S}31$, $\mathcal{S}32$, $\mathcal{S}37$, $\mathcal{S}38$, and even $\mathcal{S}43$, but the absence of other harmonics and the lower fraction of sols in which such harmonics display amplitudes over the instrumental noise level (5% for $\mathcal{S}43$) prevent us from considering those as robustly detected harmonics.

We computed periodograms for other long intervals (Supporting figures S2). $\mathcal{S}23$, absent in fig. 1, is clearly present in part of them. The periodogram for the interval spanning from sol 37 to sol 66 is particularly interesting, because C34 dust event took place during those sols, and this boosted the amplitude of solar harmonics (see subsections 2.3 and 2.4); such periodogram is much more noisy than the one shown in fig. 1, but it displays coincident peaks with much shorter periods, even beyond harmonic number 48. For the remainder of this paper we only discuss those harmonics robustly detected up to $\mathcal{S}26$.

Do these high-order solar harmonics correspond to actual thermal tides? This question can be raised from the mathematical and from the physical point of view. From the mathematical point of view, it can be argued that transient daily repeating non-tidal pressure patterns repeating at the same Local Time (LT) in consecutive sols could induce high-order solar harmonics in Fourier analysis. From the physical point of view, our analysis of high-order harmonics is limited by the fact that we rely on a single station, and therefore it is not possible to fully confirm that these harmonics are part of globally coherent tidal modes.

In order to further investigate the presence of these harmonics in the dataset, we computed the phase of the anomaly of pressure at specific ranges of the spectrum corresponding to tidal harmonics beyond $\mathcal{S}12$ and beyond $\mathcal{S}24$. The results are presented in supporting figures S1, and they reveal that wave-like patterns at those frequencies repeat similarly in different sols at most LTs (especially in some seasons, including the period of sols 490-567 corresponding to fig. 1). This shows that the measured signals inducing peaks at high-order harmonics in our periodograms are not transient daily repeating patterns, but they are present most of the time and repeat consistently in different sols, which is what we would expect if these signals corresponded to the interference of tidal harmonics. Therefore, within the limitation that we relay on a single station, our analysis is compliant with the interpretation of these high-order solar harmonics as thermal tides.

The gravitational tidal force of Phobos might produce gravitational tides on the atmosphere of Mars (as the Moon does on Earth). The orbital period of Phobos relative to a fixed point on the surface of Mars is around 40000s, but the corresponding peak is not present in our periodograms.

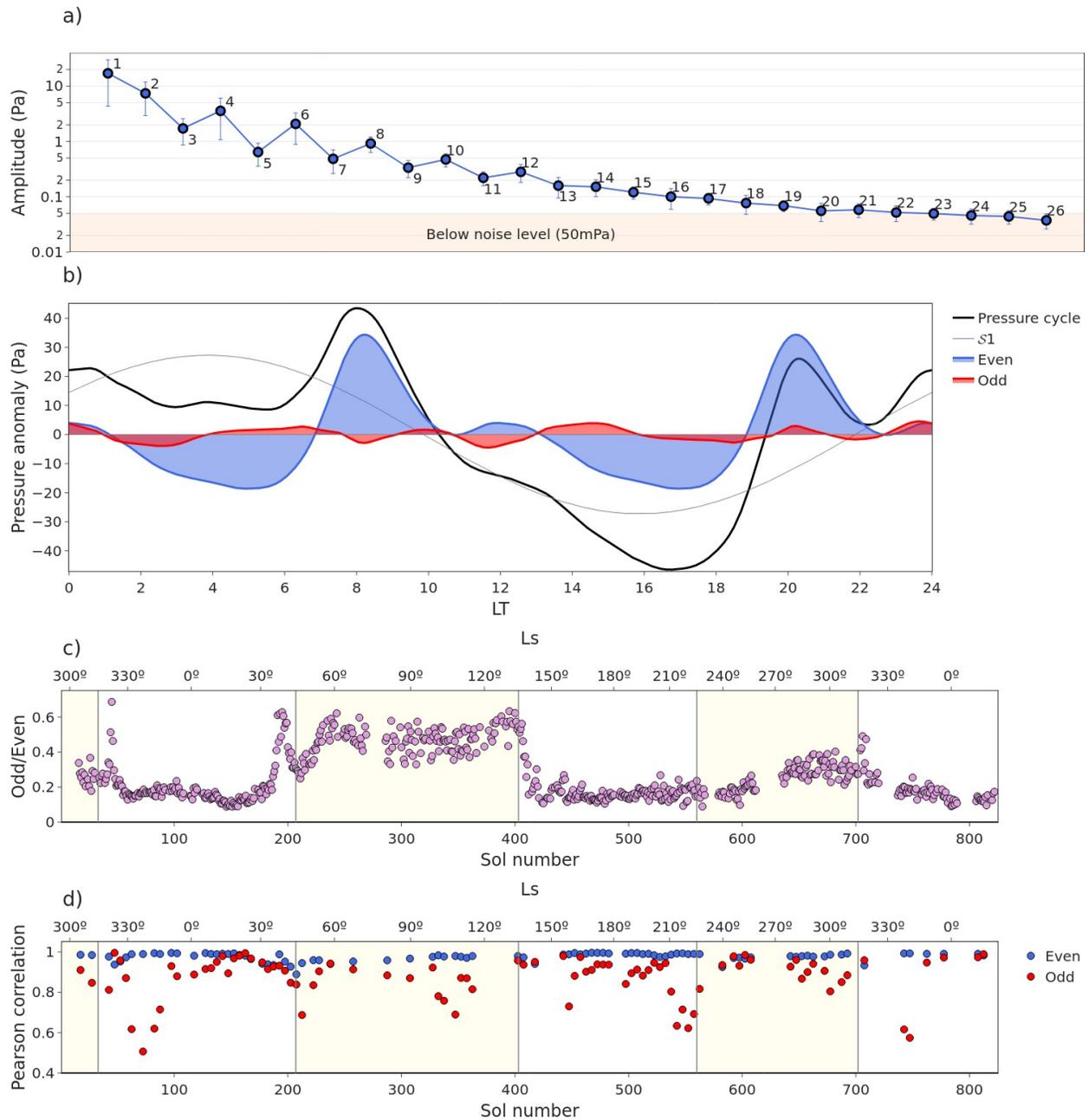

Figure 2. Differences between even and odd (excluding $S1$) harmonics. a) Daily amplitude of tidal modes averaged over a whole year, error bars represent the standard deviation. b) Example of contribution of even and odd harmonics to the diurnal cycle of pressure, corresponding to sol 129. c) Relative weight of the contribution of even and odd harmonics as function of the sol number, measured as the ratio between the filled areas in panel a. d) Pearson correlation coefficient for the linear regression of the amplitude of even and odd modes in logarithmic scale between $S2$ and $S14$. Each linear fit has been computed for the average amplitude of 5 sols, a value of 1 means perfect linear correlation. Vertical lines in panels b and c represent the limits between meteorological seasons, solstice seasons are shadowed in pale yellow.

## 2.2 Differences between even and odd harmonics

When we analyze the properties of all these tidal harmonics as computed for every sol using the Fast Fourier Transform (FFT; Cooley et al. 1965), we notice systematic differences between even and odd harmonics (excluding $S1$, which is peculiar due to its mix of migrating and non-migrating components and its vertical structure). These differences are noticeable in fig. 2a, where odd harmonics systematically exhibit smaller amplitudes between $S3$ and $S13$.

Fig. 2b is an example of the contribution of even and odd harmonics to the diurnal cycle of pressure; shadowed areas represent the pressure anomaly produced by even and odd harmonics, we find that the contribution of odd harmonics to the diurnal cycle is much smaller than that of even harmonics. Fig. 2c shows that the relative contribution of odd harmonics is larger during the northern solstice season. However, it falls abruptly with the starting of the southward equinox season (Ls 135º), when global dust content starts to rise. There is a soft increase from Ls 240º to Ls 360º. And shorter boosts take place during dust events: there are clear peaks coincident with C34, R, and C35 dust events.

Fig. 2a also shows that the averaged amplitudes of even and odd modes in logarithmic scale fall in a clear linear trend between $S2$ and $S14$. The equivalent figures for individual sols (supporting video S3) exhibit variations in the amplitudes of harmonics, and the linear trend only arises clearly when averaging some number of sols. The seasonality of this linear trend is further investigated in fig. 2d. We see that even modes follow this trend basically always, while odd modes present many exceptions.

## 2.3 Seasonal evolution of the ensemble of tidal modes

Fig. 3 is a convenient representation of the seasonal evolution of daily pressure cycle as modeled by the ensemble of tidal harmonics. It represents the variation of absolute pressure over the diurnal cycle for every sol of the dataset.

Red and blue horizontal bands are the most evident feature in this figure, these bands represent daily repeating tidal patterns and correspond to the parts of the diurnal cycle when the absolute pressure is increasing (red) or decreasing (blue). Prominent bands are present after sunrise and after sunset, they correspond to the bumps in the pressure timeseries usually observed at 8 and 20 LT, studied by Wilson et al. (2017) and Yang, Sun, et al. (2023). We note that the trends during the equinox seasons, when the sun is crossing the equator in opposite directions, are well symmetric, especially in the evening, after sunset.

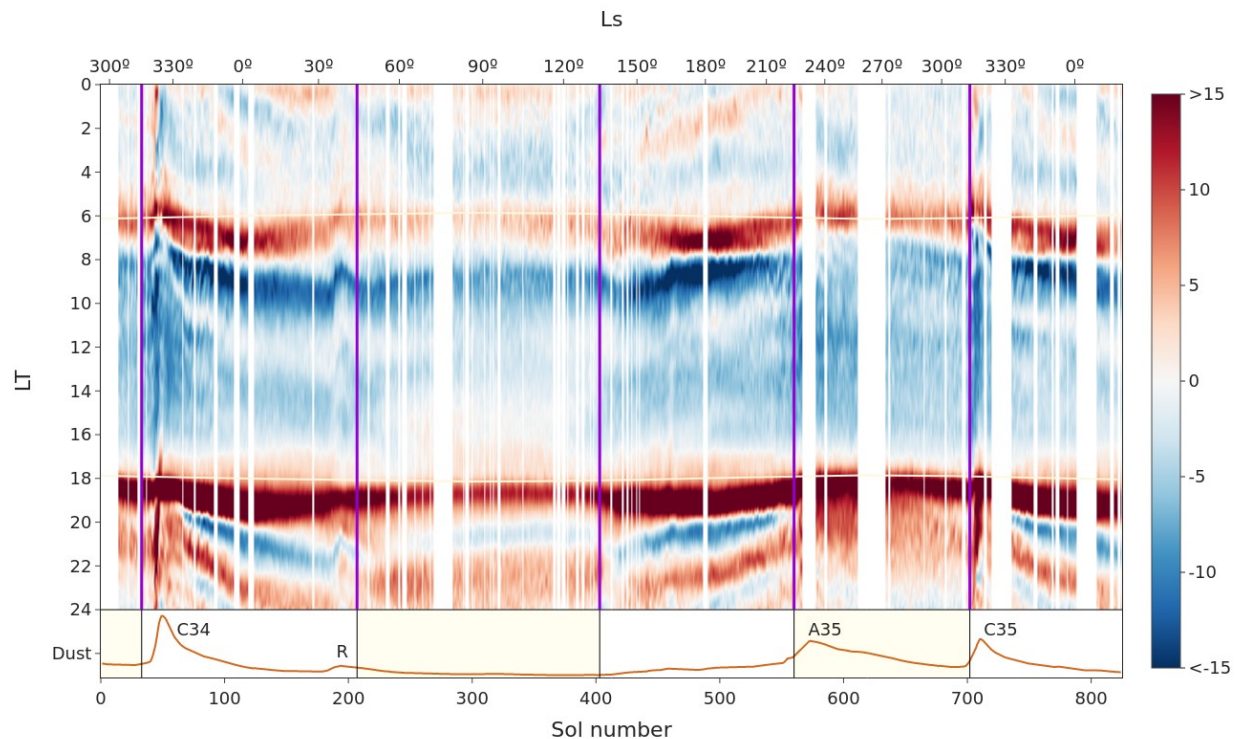

Figure 3. Climatology of the diurnal variation of pressure (dP/dt) in units of Pa/h. High frequency signals (with periods below 3700s) have been removed from the original signal before computing the derivative. Sunrise and sunset are indicated by white curves around 6 and 18 LT (Local True Solar Time). The global dust content as computed from the climatologies of Montabone et al. (2015; 2020) is represented in the bottom panel, and dust events are indicated. Seasons are delimited by vertical lines and solstice seasons are shadowed in pale yellow in the bottom (dust) panel.

In addition to the smooth seasonal evolution, other patterns at shorter temporal scales are present in this figure in coincidence with dust events C34, C35, and R. During these events, the derivatives of pressure reach higher values, corresponding to larger amplitudes of tidal modes; and the daily cycle of pressure advances, corresponding to a shift in the phases. Unfortunately, a gap without observations during peak of A35 prevents us from extracting clear conclusions about the effects of that event.

Due to their lower amplitudes (see fig. 2a), high-order harmonics are underrepresented in fig. 3. However, the phases of high-order harmonics over $\mathcal{S}12$ can be observed in supporting figures S1, and there is an obvious correlation in their seasonal evolution and that apparent here in fig. 3. This evidences that high-order harmonics evolve with a certain degree of synchronization with low-order harmonics, which is also evident in the next subsection.

**2.4 Seasonal evolution of individual tidal modes**

Each tidal harmonic is characterized by an amplitude and a phase that evolve over time. The daily values of these parameters are computed using FFT and are represented for tidal harmonics up to $\mathcal{S}12$ in fig. 4. We analyze these results in terms of seasonal variations, which are more or less coincident with the equinox and solstice seasons; and transient variations, which happen at shorter timescales of dozens of sols or less and are more likely connected to dust events or other transient meteorological phenomena. Seasonal variations are in some cases quite sudden, this is especially the case for $\mathcal{S}3$ and $\mathcal{S}4$ in the boundaries of the northern solstice season, and that is likely connected to the strong seasonal changes in the boundaries of that season previously shown in fig. 3.

Harmonics of order higher than $\mathcal{S}3$ tend to converge into overall similar seasonal trends that are in part different for even and odd harmonics. Even harmonics display larger amplitudes and delayed phases during the equinox seasons compared to the solstice seasons. Wilson et al. (2017) also found in their simulations enhanced amplitudes for $\mathcal{S}4$ and $\mathcal{S}6$ around equinoxes. Odd harmonics also display a typically delayed phase during equinox seasons, but their amplitude variations happen at shorter timescales (dozens of sols) and are in some cases more likely connected to transient phenomena (e.g. dust storms).

Harmonics $\mathcal{S}1$-$\mathcal{S}3$ exhibit their own patterns, with seasonal variations and large transient variations in response to the main dust events. The anti-correlation in the amplitudes of $\mathcal{S}3$ and $\mathcal{S}4$ observed by Guzewich et al. (2016) and Sánchez-Lavega et al. (2022) is also present in our results.

Transient dust events C34, R, A35, and C35 leave in some cases very strong footprints both in amplitudes and phases, usually in the form of peaks that only last a few sols compared to the longer length of such dust events. Amplitudes usually increase in response to dust events, but in some cases they decrease; this is especially clear in $\mathcal{S}4$ and $\mathcal{S}6$ during dust events R and C35 respectively.

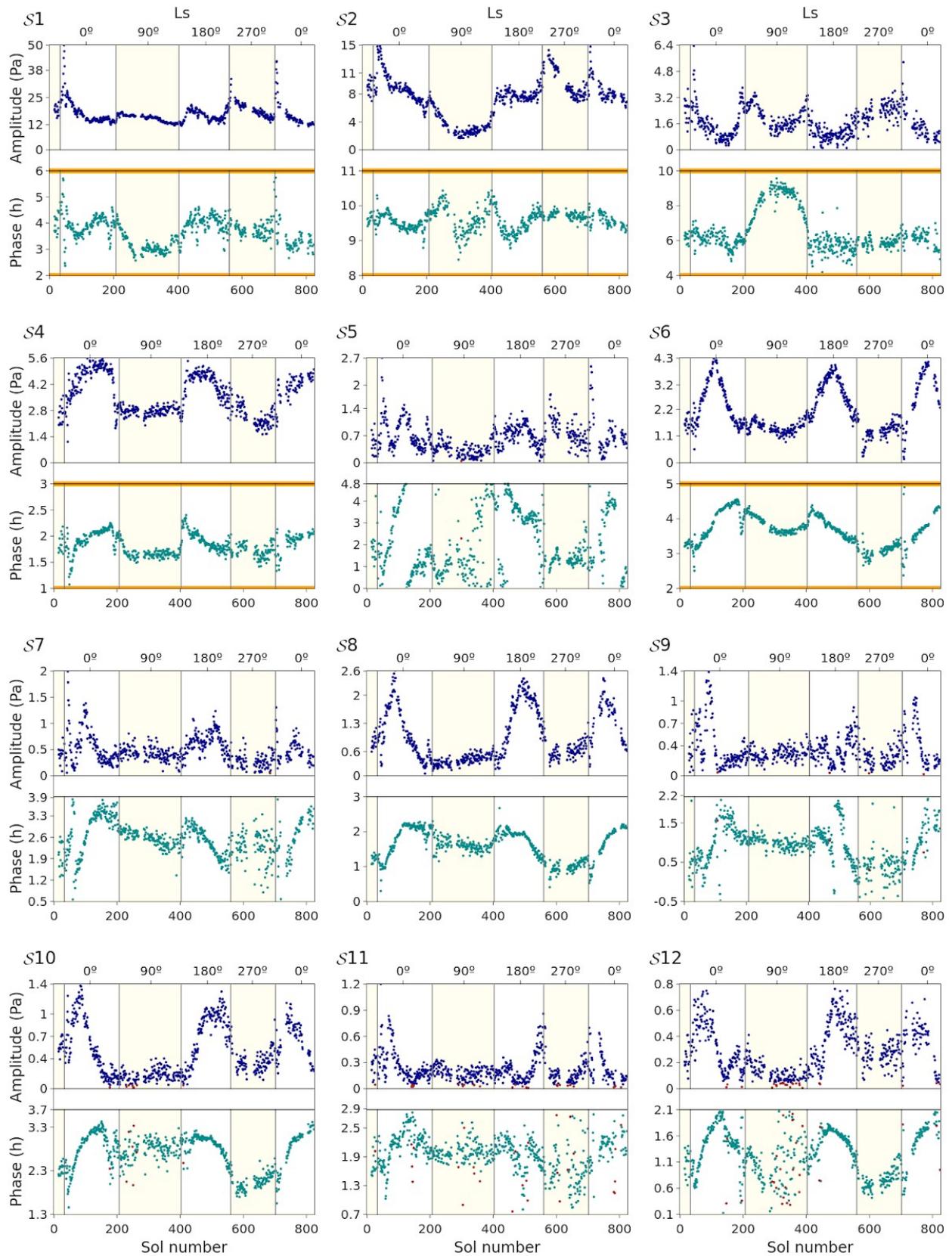

Figure 4. Amplitude (top graph in each panel; blue) and phase (bottom graph in each panel; green; expressed as LT of one maximum) of individual tidal harmonics as a function of the sol number up to $\mathcal{S}12$, which is enough to visualize the main trends found. Points below the instrumental noise level (0.05Pa) are represented in dark red. Solstice seasons are shadowed in pale yellow. In phase graphs, the vertical axis contains all the range of possible variability except for $\mathcal{S}1$, $\mathcal{S}2$, $\mathcal{S}3$, $\mathcal{S}4$ and $\mathcal{S}6$, which are very stable and their range has been constrained to appreciate better their subtle patterns within their small range of variation; this is indicated by orange horizontal lines in the top and bottom limits of the graph.

The distribution of phases exhibits different levels of randomness that are different in different tidal harmonics and seasons. It tends to be more random in odd and higher order harmonics, and in solstice seasons (northern solstice: $\mathcal{S}5$, $\mathcal{S}10$, $\mathcal{S}12$; southern solstice: $\mathcal{S}7$, $\mathcal{S}9$, $\mathcal{S}11$). The equivalent graphs for harmonics higher than $\mathcal{S}12$ are available in the supporting figure S4, and they show that phases look more and more random for those higher order harmonics

## 3 Summary and perspectives

Our analysis supports that high-order tidal harmonics (even beyond $\mathcal{S}24$; subsection 2.1) are present in the dataset acquired by the PS on Insight. The finding of a large number of harmonics and the excellent temporal coverage of this dataset have enabled us to explore the seasonal evolution of: I) The differences between even and odd harmonics (subsections 2.2 and 2.4), II) The ensemble of thermal tides in the daily cycle of pressure (subsection 2.3), III) The individual tidal harmonics (subsection 2.4).

We analyzed seasonal variations in terms of seasons centered on equinoxes and solstices, and our results adapt very well to them, with $\mathcal{S}3$ and $\mathcal{S}4$ experiencing sudden changes exactly in the limits of such seasons. The low thermal inertia of the Martian atmosphere makes it respond more quickly to changes in insolation, and that is likely the reason why this definition of seasons centered in equinoxes and solstices adapts so good to certain observations of the Martian atmosphere.

The average amplitude of tidal harmonics tends to fall exponentially between $\mathcal{S}2$ and $\mathcal{S}14$, with different rates for even and odd harmonics (fig. 2a); the latter exhibit systematically smaller amplitudes. These trends undergo a seasonal evolution, possibly connected to the evolution of thermal forcing depending on insolation and aerosols distribution. Theoretical and modeling studies are needed to further investigate the processes leading to these trends. High order and odd harmonics tend to exhibit more random variations in their phase and amplitude, their smaller amplitudes probably make them more susceptible to slight changes in the atmosphere, and therefore they might respond more easily to the presence of aerosols. The seasonally larger amplitude of even harmonics around equinoxes matches with the finding of the same trend for $S4$ and $S6$ in simulations (Wilson et al., 2017).

The existence of high-order harmonics on Mars that repeat similarly in different sols, probably with some degree of interannual repeatability, could have consequences on the mesoscale and the microscale in the form of daily repeating patterns of subtle variations of meteorological variables that in some cases could trigger daily repeating phenomena. This is especially the case because a large number of low amplitude harmonics with similar frequencies can transiently interfere constructively to produce signals with larger amplitudes (as also suggested for Earth by Hedlin et al., 2018), and such signals could even be confused with gravity waves (a possibility that we will explore as part of a separate work).

In any case, the present paper opens a new window for theoretical works and modeling to explore features reported here (e.g. differences between even and odd harmonics), check whether models reproduce the observed behavior of high-order harmonics, use such models to learn more about these harmonics, determine the contribution of different mechanisms to the different tidal harmonics (e.g. Geißler et al., 2020), and explore the possible consequences and effects of high-order tidal harmonics, on Mars and maybe on other planets as well. When it comes to observations, many aspects of the Insight dataset remain to be explored, and other missions currently operating on Mars can try to find high-order harmonics, and the coordinated analysis of data from those missions can contribute to investigate their global structure. This work enhances the interest to include pressure sensors with improved accuracy in future landers, and to deploy a network to make simultaneous observations.

## Acknowledgments

This is InSight contribution number 336. The authors acknowledge the funding support provided by Agence Nationale de la Recherche (ANR-19-CE31-0008-08 MAGIS) and CNES. All co-authors acknowledge NASA, Centre National d'Études Spatiales (CNES) and its partner agencies and institutions (UKSA, SSO, DLR, JPL, IPGP-CNRS, ETHZ, IC, and MPS-MPG), and the flight operations team at JPL, CAB, SISMOC, MSDS, IRIS-DMC, and PDS for providing InSight data. The members of the InSight engineering and operations teams made the InSight mission possible and their hard work and dedication is acknowledged here.

We thank Lucas Lange for his useful comments. And we thank John Wilson and an anonymous reviewer for their excellent work reviewing this paper, which largely contributed to improve it.

JHB expresses his support to UN Secretary General António Guterres in his calls to stop violations of Human Rights and International Law, and to act against climate change. JHB expresses his support to scientist colleagues on trial and in jail across the world for their non-violent actions against climate inaction. https://scientistrebellion.org/scientists-on-trial/

**Open Research**

The derived dataset of continuous timeseries produced for this work can be accessed at Hernández-Bernal et al. (2023). Data resulting from this work and used to generate figures can be accessed in the repository Hernández-Bernal et al. (2024).

All InSight data used in this study are publicly available in the Planetary Data System (PDS; Banfield, 2020). The derived dataset of intervals with continuous coverage can be accessed in

Most recent dust climatologies can be obtained at Montabone et al. (2022).

# Supporting Information

**Contents of this file**

    Text S1-S3

    Tables S1-S3

    Figures S1 to S4

**Additional Supporting Information (Files uploaded separately, they can be found at https://doi.org/10.1029/2023GL107674 in the bottom of the page)**

    Captions for datasets S1-S4

    Caption for videos S1-S3

**Introduction**

This supporting material includes:

- Text S1. Notes on Methodology
- Figures S1. Climatology of phase of pressure anomalies at different parts of the spectrum.
- Caption for videos S1 and S2. Showing the coherence of pressure anomalies trough different sols.
- Table S1. Table containing intervals with continuous measurements without gaps longer than 100s.
- Table S2. List of gaps in intervals fro Table S1.
- Table S3. List of full sols with continuous coverage as derived from table S1.
- Figures S2. Equivalent to fig. 1.
- Text S2. About the computation of average and noise level in fig. 1.
- Figures S3. Equivalent to fig. 2a, for meteorological seasons.
- Caption for video S3. Equivalent to fig. 2a, for each sol.
- Figure S4. Analogue to fig. 4 for harmonics 13-27.
- Text S3. Describing the format of the datasets.
- Captions for datasets S1-S4. With all data used to create figures.

**Text S1. Notes on Methodology.**

We make two different types of analysis: periodograms to explore the existence of tidal modes in the data; and Fast Fourier Transforms (FFT; Cooley et al. 1965) to find the specific parameter of such tidal modes for each sol. We use the implementations made by Harris et al. (2020), both for periodograms and FFT.

The quality of periodograms is better when they are generated from very long timeseries. We make a list of intervals containing continuous measurements over dozens of sols without gaps longer than 70s (see tables table S1 and S2 for details), and we use the longest intervals for our periodograms. Gaps smaller than 70 s are filled with linear interpolations.

When it comes to determine the parameters of each tidal mode for each sol, we use continuous timeseries covering the whole sol (using Local True Solar Time to split sols), this leaves us with 630 full sols between sol 15 and sol 824, both included. This list of full sols is provided in table S3.

**Figures S1. Climatology of phase of pressure anomalies at different parts of the spectrum.** Each point in these figures represents the sign of the pressure anomaly after the application of a bandpass filter. Red corresponds to positive anomalies, blue corresponds to negative anomalies. The horizontal bands represent LTs when the phase of the bandpassed signal is similar in consecutive sols.

For clarity, we show an additional figure showing the bandpassed pressure timeseries in sol 115 and the corresponding red (positive) and blue (negative) colors assigned to pressure anomalies after such bandpass. Animated versions of these subfigures are available in supporting videos S2 and S3.

**Bandpass at 3700-7400s (1-2 Martian hours, $\mathcal{S}12$ to $\mathcal{S}24$)**

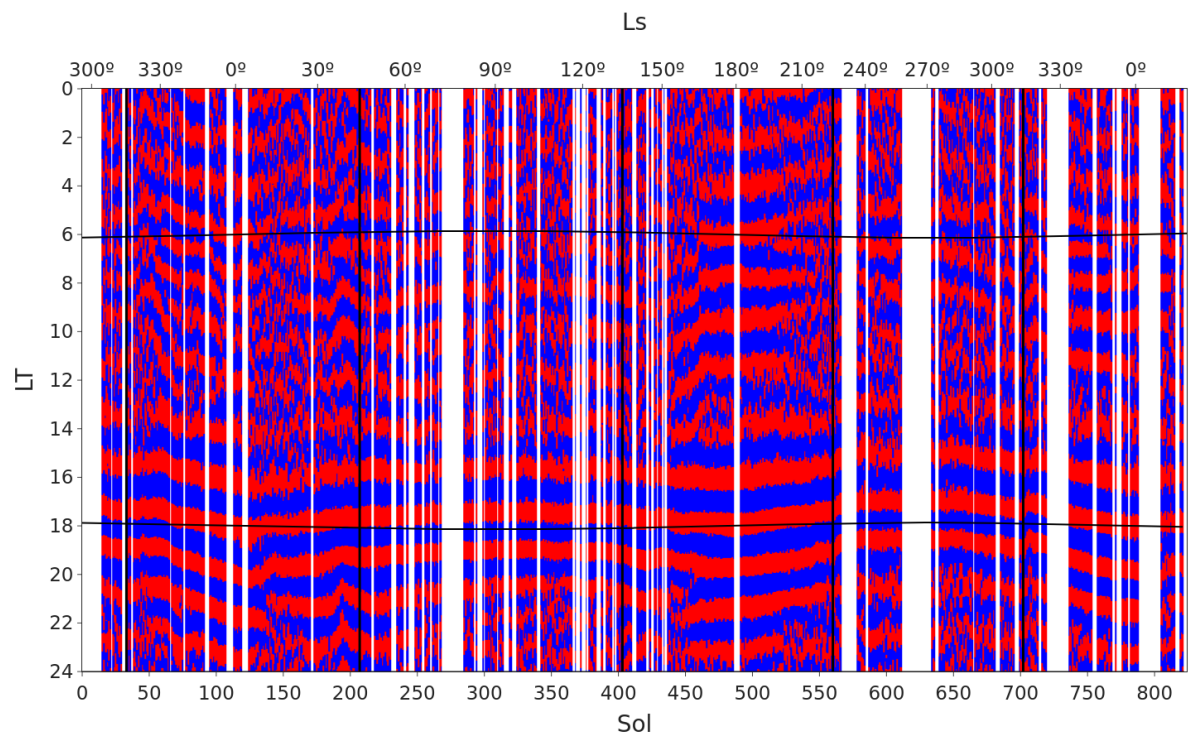

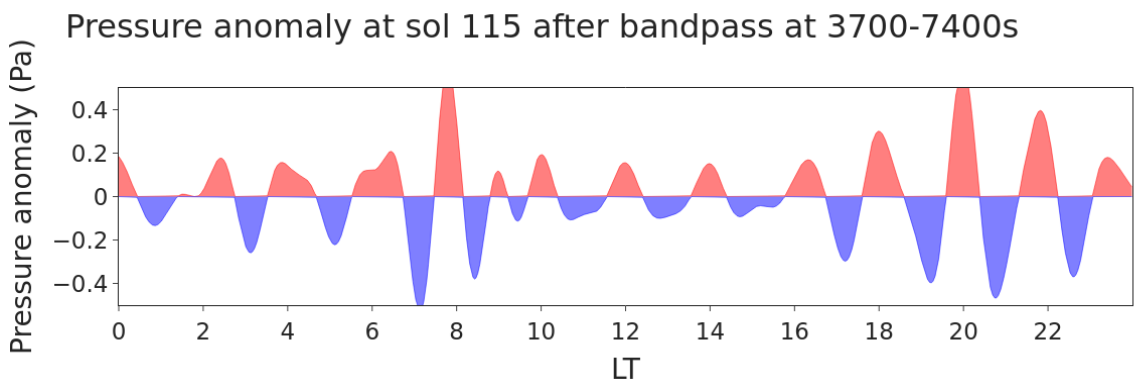

**Bandpass at 1800-3700s (0.5-1 Martian hours, $\mathcal{S}24$ to $\mathcal{S}48$)**

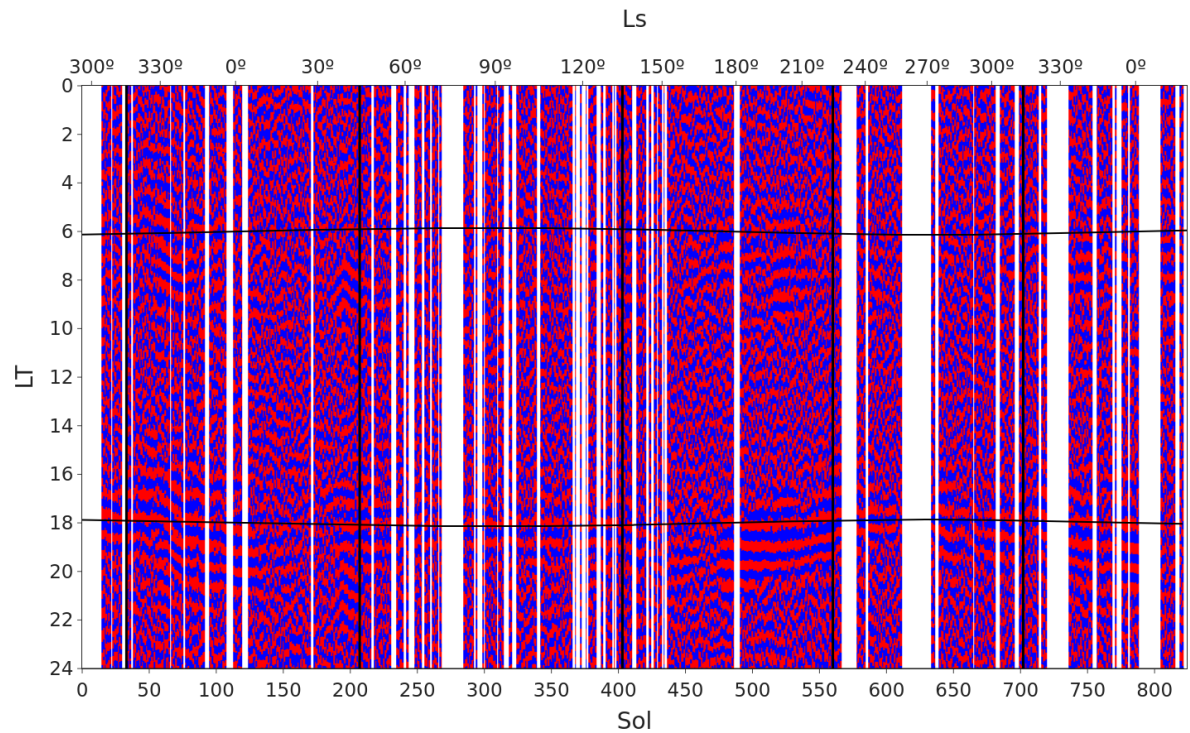

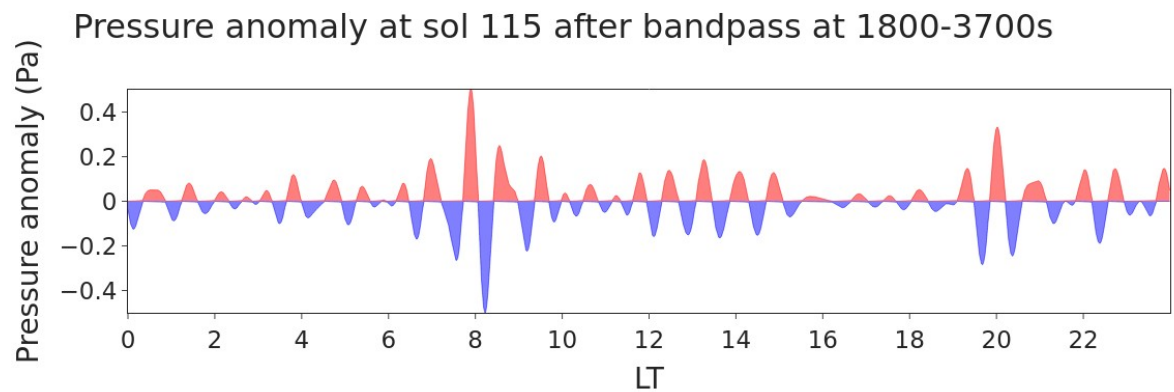

**Caption for videos S1 and S2.** These videos show the pressure anomalies for each sol after bandpass. The color of the shadowing indicates the positive (red) or negative (blue) anomaly, and corresponds to the colors assigned in supporting figures S1.

**Table S1. List of intervals without gaps longer than 100s.** For each interval, the starting (sol1) and ending (sol2) sols are indicated. LT1 and LT2 refer to the local time of the starting and ending of the interval within the sol.

| ID# | Sol 1 | LT1 | Sol2 | LT2 | ID# | Sol 1 | LT1 | Sol2 | LT2 |
|---|---|---|---|---|---|---|---|---|---|
| 0 | 4 | 6.7 | 4 | 6.8 | 50 | 377 | 7.7 | 384 | 10.7 |
| 1 | 4 | 11.1 | 4 | 11.3 | 51 | 384 | 15.7 | 385 | 19.4 |
| 2 | 5 | 13 | 5 | 17.2 | 52 | 386 | 2.6 | 389 | 7.6 |
| 3 | 10 | 14.5 | 10 | 14.9 | 53 | 389 | 8.1 | 390 | 16.5 |
| 4 | 14 | 11.5 | 22 | 15.4 | 54 | 390 | 17.4 | 390 | 17.5 |
| 5 | 22 | 19.5 | 30 | 5.1 | 55 | 390 | 18.1 | 396 | 6.6 |
| 6 | 32 | 8.3 | 37 | 12.3 | 56 | 396 | 7.2 | 396 | 15.9 |
| 7 | 37 | 13.8 | 66 | 9.4 | 57 | 396 | 16.2 | 398 | 11 |
| 8 | 66 | 11.6 | 76 | 9.7 | 58 | 398 | 11.6 | 412 | 0.1 |
| 9 | 76 | 11.9 | 92 | 17.9 | 59 | 412 | 1.2 | 412 | 21.1 |
| 10 | 94 | 6.4 | 108 | 17.8 | 60 | 413 | 0.7 | 421 | 19.4 |
| 11 | 112 | 9 | 120 | 6 | 61 | 422 | 2.7 | 425 | 9.7 |
| 12 | 123 | 6.9 | 171 | 17.7 | 62 | 425 | 10.3 | 425 | 20.8 |
| 13 | 172 | 2.3 | 216 | 19.1 | 63 | 425 | 22 | 426 | 7.1 |
| 14 | 217 | 12.8 | 231 | 15 | 64 | 426 | 7.7 | 429 | 8.3 |
| 15 | 232 | 15.6 | 233 | 20.2 | 65 | 429 | 8.9 | 433 | 11.7 |
| 16 | 233 | 21 | 241 | 1.3 | 66 | 433 | 12.4 | 435 | 16.9 |
| 17 | 241 | 1.9 | 244 | 14.3 | 67 | 435 | 21.1 | 487 | 5.2 |
| 18 | 244 | 15 | 245 | 10.5 | 68 | 487 | 6.7 | 488 | 10.7 |
| 19 | 245 | 11.1 | 246 | 16 | 69 | 488 | 16.4 | 489 | 8.9 |
| 20 | 247 | 0.6 | 254 | 19.5 | 70 | 489 | 9.6 | 490 | 14.8 |
| 21 | 254 | 21 | 260 | 11.7 | 71 | 490 | 15.5 | 567 | 9.9 |

| | | | | | | | | |
|---|---|---|---|---|---|---|---|---|
| 22 | 260 | 15.3 | 266 | 11.5 | 72 | 577 | 9.2 | 585 | 17.9 |
| 23 | 266 | 19.5 | 269 | 5.6 | 73 | 586 | 0 | 612 | 5.7 |
| 24 | 284 | 15.2 | 293 | 14.7 | 74 | 633 | 8 | 637 | 12.5 |
| 25 | 293 | 15.6 | 293 | 15.6 | 75 | 637 | 21.7 | 665 | 6.8 |
| 26 | 293 | 16.2 | 295 | 4.3 | 76 | 665 | 12.6 | 682 | 16.3 |
| 27 | 295 | 5.5 | 295 | 23.1 | 77 | 683 | 22.4 | 696 | 17.6 |
| 28 | 296 | 0.2 | 297 | 19.8 | 78 | 697 | 6 | 698 | 12.4 |
| 29 | 297 | 20.1 | 300 | 16 | 79 | 698 | 13.4 | 702 | 5.4 |
| 30 | 300 | 16.8 | 310 | 14.8 | 80 | 702 | 17.8 | 714 | 14.7 |
| 31 | 310 | 15.3 | 315 | 14.5 | 81 | 715 | 1.6 | 720 | 8.1 |
| 32 | 315 | 15.2 | 317 | 1.9 | 82 | 724 | 19.9 | 725 | 12.5 |
| 33 | 317 | 2.5 | 317 | 18.4 | 83 | 726 | 21.8 | 727 | 12.9 |
| 34 | 317 | 21.7 | 321 | 21.7 | 84 | 727 | 21.3 | 728 | 12.9 |
| 35 | 321 | 22.3 | 322 | 12.1 | 85 | 729 | 21.3 | 730 | 12.9 |
| 36 | 322 | 13.6 | 322 | 20.2 | 86 | 735 | 16.9 | 754 | 7.6 |
| 37 | 322 | 21 | 341 | 1.7 | 87 | 756 | 1.5 | 769 | 13 |
| 38 | 341 | 2.6 | 341 | 2.7 | 88 | 769 | 21.3 | 772 | 14.3 |
| 39 | 341 | 3.4 | 366 | 17.4 | 89 | 772 | 18.6 | 773 | 12.4 |
| 40 | 367 | 2 | 369 | 9.3 | 90 | 773 | 15 | 774 | 10.6 |
| 41 | 369 | 10.2 | 369 | 10.2 | 91 | 774 | 22.1 | 781 | 5.2 |
| 42 | 369 | 10.8 | 370 | 5.3 | 92 | 781 | 6.9 | 789 | 6 |
| 43 | 370 | 6.8 | 370 | 11.4 | 93 | 793 | 18.9 | 794 | 7.1 |
| 44 | 370 | 13.9 | 371 | 8 | 94 | 802 | 18.5 | 803 | 16.2 |
| 45 | 371 | 8.9 | 371 | 8.9 | 95 | 804 | 3 | 816 | 7.7 |
| 46 | 371 | 9.4 | 373 | 23.3 | 96 | 818 | 3.1 | 822 | 8 |

| 47 | 374 | 0.6 | 375 | 3.3 | 97 | 822 | 16.6 | 823 | 9 |
| 48 | 375 | 3.8 | 375 | 6.6 | 98 | 823 | 21.3 | 825 | 5.6 |
| 49 | 375 | 7.1 | 377 | 6.5 | | | | | |

**Table S2. List of gaps with length between 5 and 70 s in intervals of table S1.** Gaps shorter than 70 s were linearly interpolated. Gaps longer than 70 s split intervals in table S1.

| Sol # | Nº of gaps | Lengths of gaps (s) | Sol # | Nº of gaps | Lengths of gaps (s) |
|---|---|---|---|---|---|
| 22 | 1 | 16 | 371 | 2 | 5 |
| 48 | 1 | 44 | 374 | 1 | 5 |
| 67 | 1 | 44 | 375 | 6 | 5, 69 |
| 84 | 1 | 44 | 389 | 2 | 5 |
| 113 | 1 | 44 | 390 | 3 | 5 |
| 143 | 1 | 44 | 396 | 6 | 5 |
| 182 | 1 | 6 | 398 | 3 | 5 |
| 204 | 1 | 7 | 412 | 1 | 5 |
| 230 | 1 | 6 | 413 | 4 | 5 |
| 233 | 5 | 5 | 416 | 1 | 7 |
| 241 | 3 | 5 | 425 | 4 | 5 |
| 244 | 3 | 5 | 426 | 3 | 5 |
| 245 | 3 | 5 | 429 | 3 | 5 |
| 247 | 3 | 5 | 433 | 2 | 5 |
| 254 | 1 | 5 | 434 | 1 | 7 |
| 256 | 1 | 7 | 467 | 1 | 7 |
| 284 | 2 | 47, 53 | 472 | 1 | 7 |
| 292 | 1 | 6 | 475 | 1 | 6 |
| 297 | 1 | 5 | 482 | 1 | 6 |
| 300 | 3 | 5 | 487 | 1 | 5 |
| 310 | 3 | 5 | 489 | 5 | 5, 6 |

| 315 | 2 | 5 | 490 | 4 | 5 |
|-----|---|---|-----|---|---|
| 317 | 5 | 5 | 516 | 1 | 7 |
| 321 | 2 | 5 | 521 | 1 | 7 |
| 322 | 4 | 5 | 523 | 1 | 7 |
| 340 | 1 | 7 | 533 | 1 | 7 |
| 341 | 3 | 5 | 534 | 1 | 7 |
| 365 | 1 | 7 | 740 | 1 | 7 |
| 369 | 2 | 5 | 822 | 1 | 12 |
| 370 | 2 | 5 |     |   |    |

**Table S3. List of full sols with continuous coverage as derived from table S1.**

| | | | | | | | | | | | | | | |
|---|---|---|---|---|---|---|---|---|---|---|---|---|---|---|
| 15 | 16 | 17 | 18 | 19 | 20 | 21 | 23 | 24 | 25 | 26 | 27 | 28 | 29 | 33 |
| 34 | 35 | 36 | 38 | 39 | 40 | 41 | 42 | 43 | 44 | 45 | 46 | 47 | 48 | 49 |
| 50 | 51 | 52 | 53 | 54 | 55 | 56 | 57 | 58 | 59 | 60 | 61 | 62 | 63 | 64 |
| 65 | 67 | 68 | 69 | 70 | 71 | 72 | 73 | 74 | 75 | 77 | 78 | 79 | 80 | 81 |
| 82 | 83 | 84 | 85 | 86 | 87 | 88 | 89 | 90 | 91 | 95 | 96 | 97 | 98 | 99 |
| 100 | 101 | 102 | 103 | 104 | 105 | 106 | 107 | 113 | 114 | 115 | 116 | 117 | 118 | 119 |
| 124 | 125 | 126 | 127 | 128 | 129 | 130 | 131 | 132 | 133 | 134 | 135 | 136 | 137 | 138 |
| 139 | 140 | 141 | 142 | 143 | 144 | 145 | 146 | 147 | 148 | 149 | 150 | 151 | 152 | 153 |
| 154 | 155 | 156 | 157 | 158 | 159 | 160 | 161 | 162 | 163 | 164 | 165 | 166 | 167 | 168 |
| 169 | 170 | 173 | 174 | 175 | 176 | 177 | 178 | 179 | 180 | 181 | 182 | 183 | 184 | 185 |
| 186 | 187 | 188 | 189 | 190 | 191 | 192 | 193 | 194 | 195 | 196 | 197 | 198 | 199 | 200 |
| 201 | 202 | 203 | 204 | 205 | 206 | 207 | 208 | 209 | 210 | 211 | 212 | 213 | 214 | 215 |
| 218 | 219 | 220 | 221 | 222 | 223 | 224 | 225 | 226 | 227 | 228 | 229 | 230 | 234 | 235 |
| 236 | 237 | 238 | 239 | 240 | 242 | 243 | 248 | 249 | 250 | 251 | 252 | 253 | 255 | 256 |
| 257 | 258 | 259 | 261 | 262 | 263 | 264 | 265 | 267 | 268 | 285 | 286 | 287 | 288 | 289 |
| 290 | 291 | 292 | 294 | 298 | 299 | 301 | 302 | 303 | 304 | 305 | 306 | 307 | 308 | 309 |
| 311 | 312 | 313 | 314 | 316 | 318 | 319 | 320 | 323 | 324 | 325 | 326 | 327 | 328 | 329 |
| 330 | 331 | 332 | 333 | 334 | 335 | 336 | 337 | 338 | 339 | 340 | 342 | 343 | 344 | 345 |
| 346 | 347 | 348 | 349 | 350 | 351 | 352 | 353 | 354 | 355 | 356 | 357 | 358 | 359 | 360 |
| 361 | 362 | 363 | 364 | 365 | 368 | 372 | 376 | 378 | 379 | 380 | 381 | 382 | 383 | 387 |
| 388 | 391 | 392 | 393 | 394 | 395 | 397 | 399 | 400 | 401 | 402 | 403 | 404 | 405 | 406 |
| 407 | 408 | 409 | 410 | 411 | 414 | 415 | 416 | 417 | 418 | 419 | 420 | 423 | 424 | 427 |
| 428 | 430 | 431 | 432 | 434 | 436 | 437 | 438 | 439 | 440 | 441 | 442 | 443 | 444 | 445 |

| 446 | 447 | 448 | 449 | 450 | 451 | 452 | 453 | 454 | 455 | 456 | 457 | 458 | 459 | 460 |
|---|---|---|---|---|---|---|---|---|---|---|---|---|---|---|
| 461 | 462 | 463 | 464 | 465 | 466 | 467 | 468 | 469 | 470 | 471 | 472 | 473 | 474 | 475 |
| 476 | 477 | 478 | 479 | 480 | 481 | 482 | 483 | 484 | 485 | 486 | 491 | 492 | 493 | 494 |
| 495 | 496 | 497 | 498 | 499 | 500 | 501 | 502 | 503 | 504 | 505 | 506 | 507 | 508 | 509 |
| 510 | 511 | 512 | 513 | 514 | 515 | 516 | 517 | 518 | 519 | 520 | 521 | 522 | 523 | 524 |
| 525 | 526 | 527 | 528 | 529 | 530 | 531 | 532 | 533 | 534 | 535 | 536 | 537 | 538 | 539 |
| 540 | 541 | 542 | 543 | 544 | 545 | 546 | 547 | 548 | 549 | 550 | 551 | 552 | 553 | 554 |
| 555 | 556 | 557 | 558 | 559 | 560 | 561 | 562 | 563 | 564 | 565 | 566 | 578 | 579 | 580 |
| 581 | 582 | 583 | 584 | 587 | 588 | 589 | 590 | 591 | 592 | 593 | 594 | 595 | 596 | 597 |
| 598 | 599 | 600 | 601 | 602 | 603 | 604 | 605 | 606 | 607 | 608 | 609 | 610 | 611 | 634 |
| 635 | 636 | 638 | 639 | 640 | 641 | 642 | 643 | 644 | 645 | 646 | 647 | 648 | 649 | 650 |
| 651 | 652 | 653 | 654 | 655 | 656 | 657 | 658 | 659 | 660 | 661 | 662 | 663 | 664 | 666 |
| 667 | 668 | 669 | 670 | 671 | 672 | 673 | 674 | 675 | 676 | 677 | 678 | 679 | 680 | 681 |
| 684 | 685 | 686 | 687 | 688 | 689 | 690 | 691 | 692 | 693 | 694 | 695 | 699 | 700 | 701 |
| 703 | 704 | 705 | 706 | 707 | 708 | 709 | 710 | 711 | 712 | 713 | 716 | 717 | 718 | 719 |
| 736 | 737 | 738 | 739 | 740 | 741 | 742 | 743 | 744 | 745 | 746 | 747 | 748 | 749 | 750 |
| 751 | 752 | 753 | 757 | 758 | 759 | 760 | 761 | 762 | 763 | 764 | 765 | 766 | 767 | 768 |
| 770 | 771 | 775 | 776 | 777 | 778 | 779 | 780 | 782 | 783 | 784 | 785 | 786 | 787 | 788 |
| 805 | 806 | 807 | 808 | 809 | 810 | 811 | 812 | 813 | 814 | 815 | 819 | 820 | 821 | 824 |

**Figures S2. Analogue to Fig. 1, for all intervals longer than 10 sols.** Titles of figures are in the format: *Interval ID#. Sols [Sol1 from table S1]-[Sol2 from table S1] (Number of sols)*. Note that the noise level in this periodograms is in general higher than in fig. 1, and therefore we used less strict criteria to compute the noise level. See supporting text S2 for details.

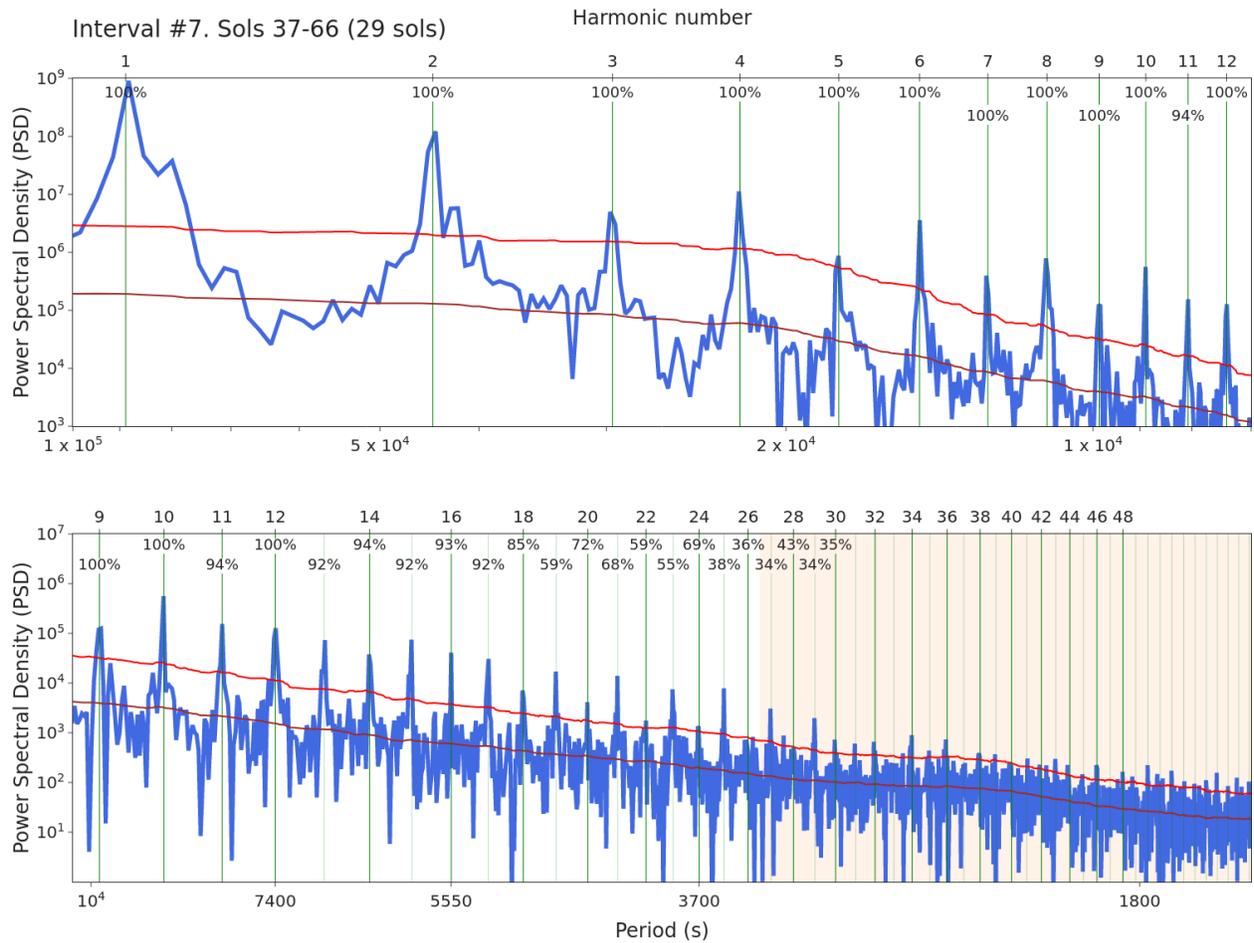

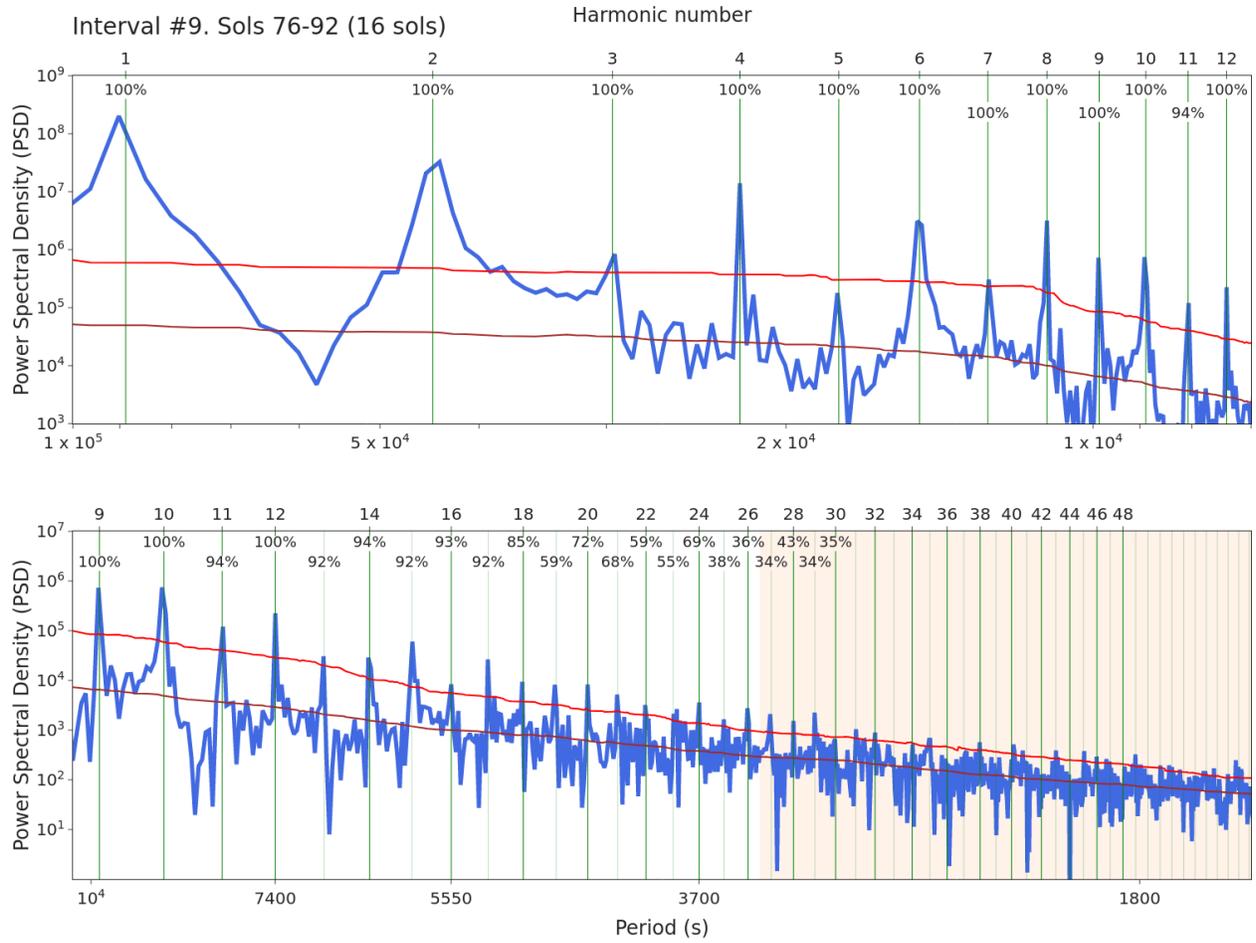

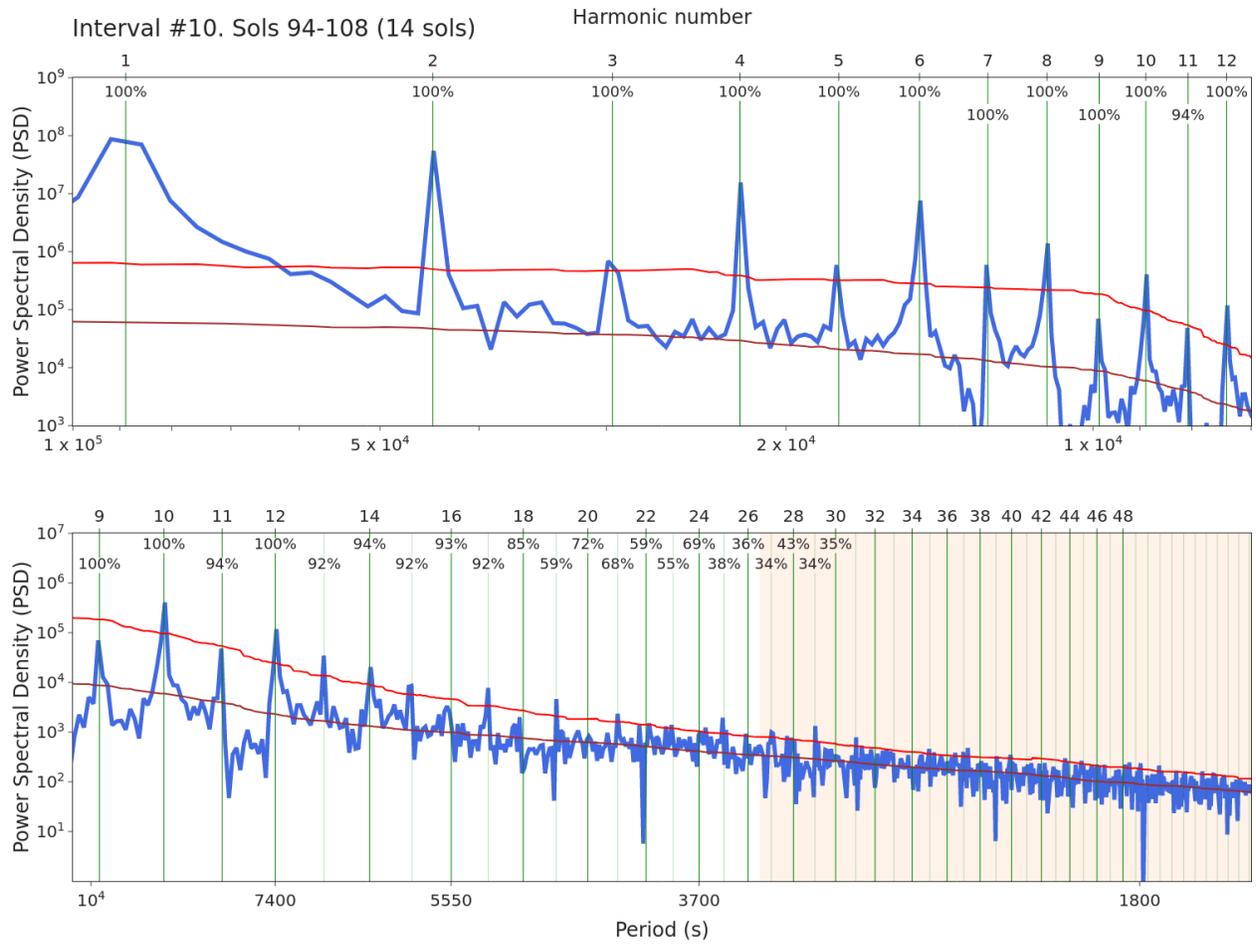

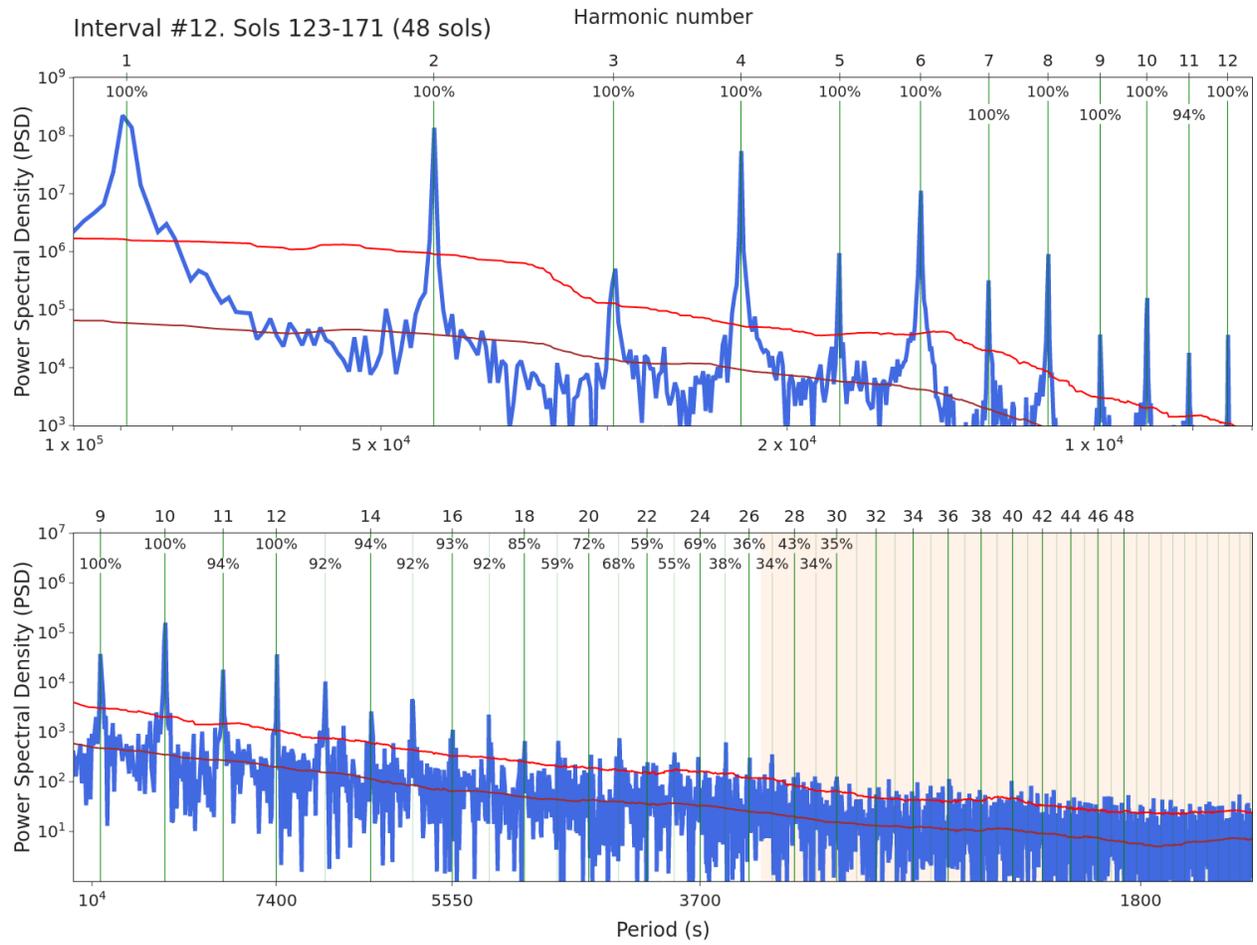

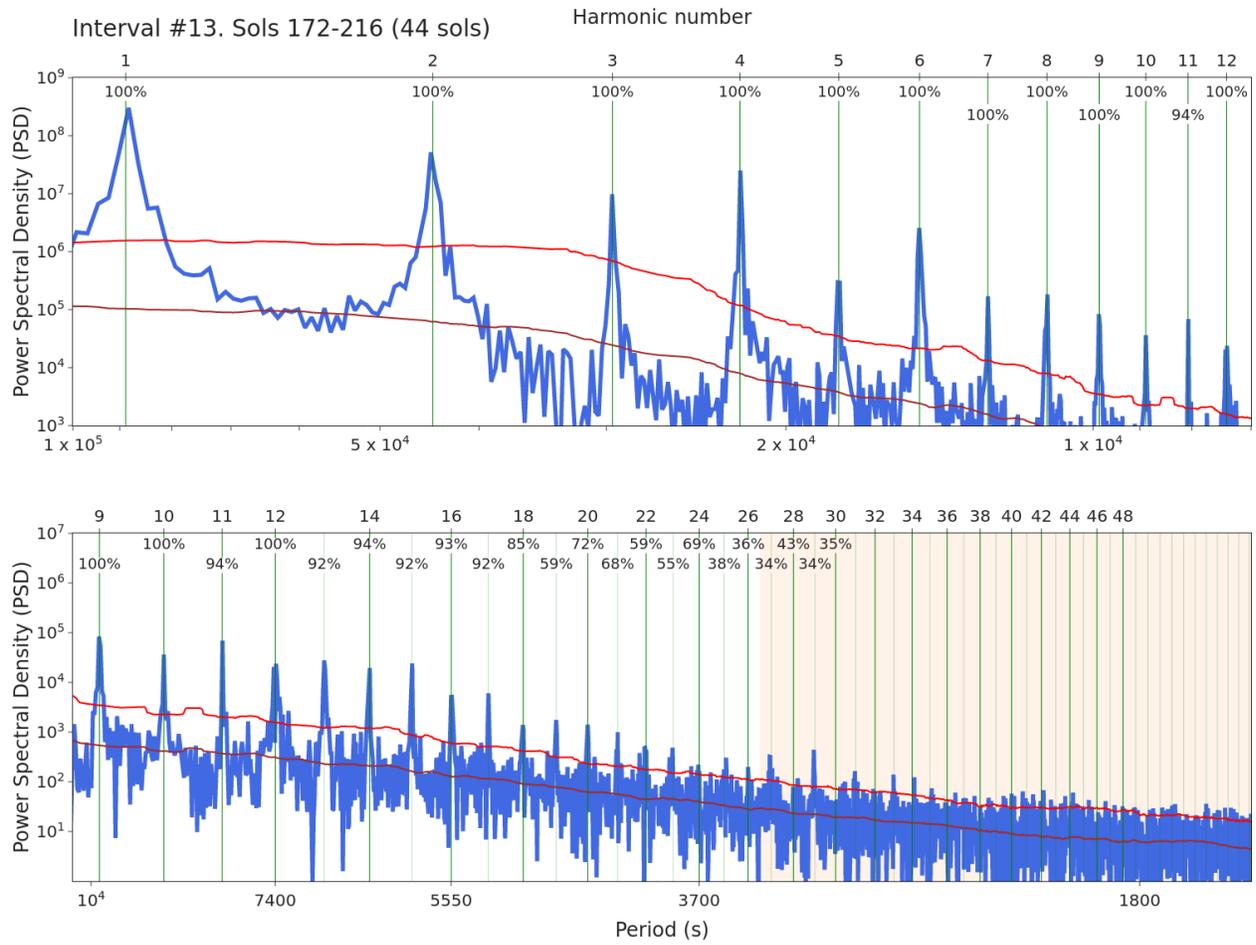

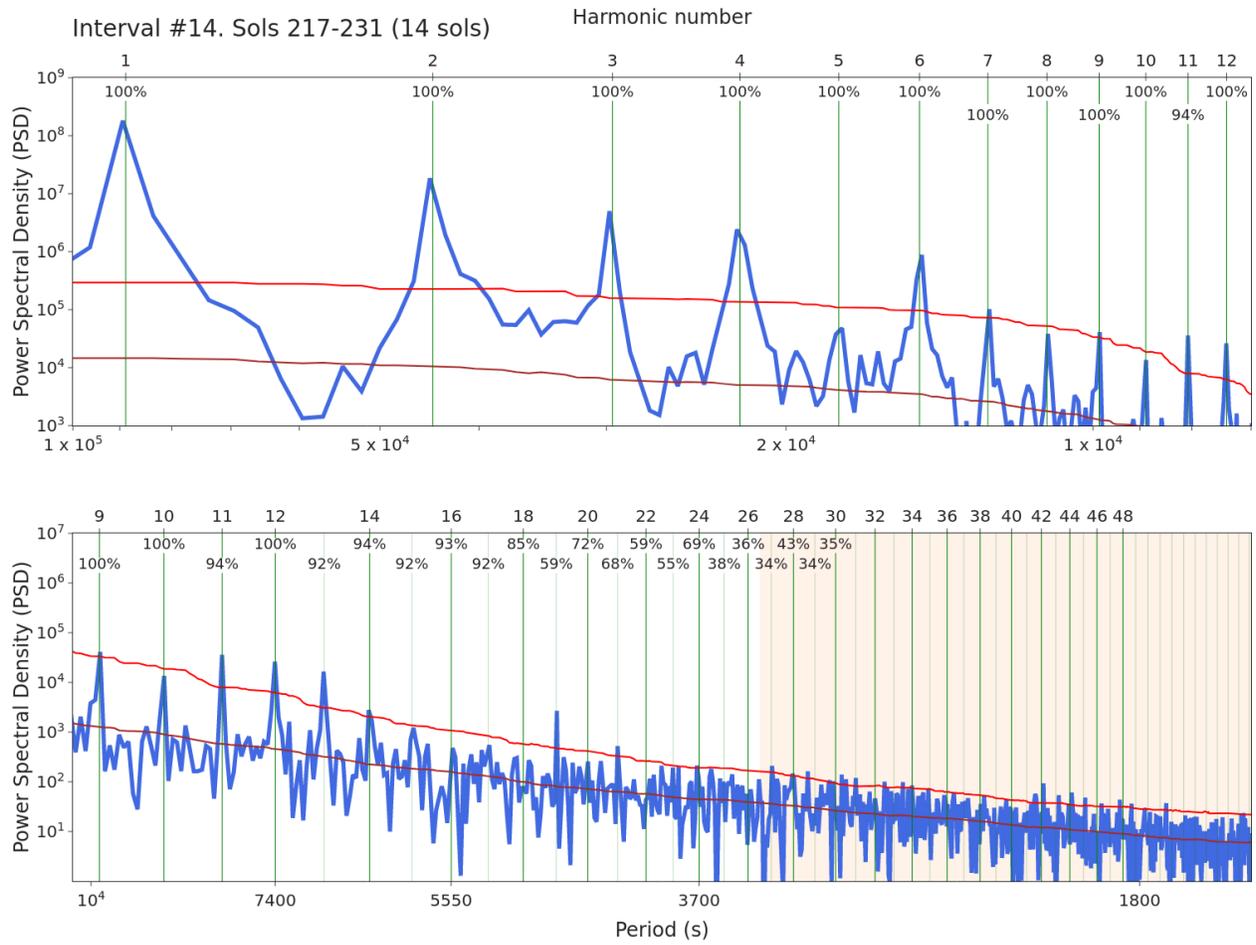

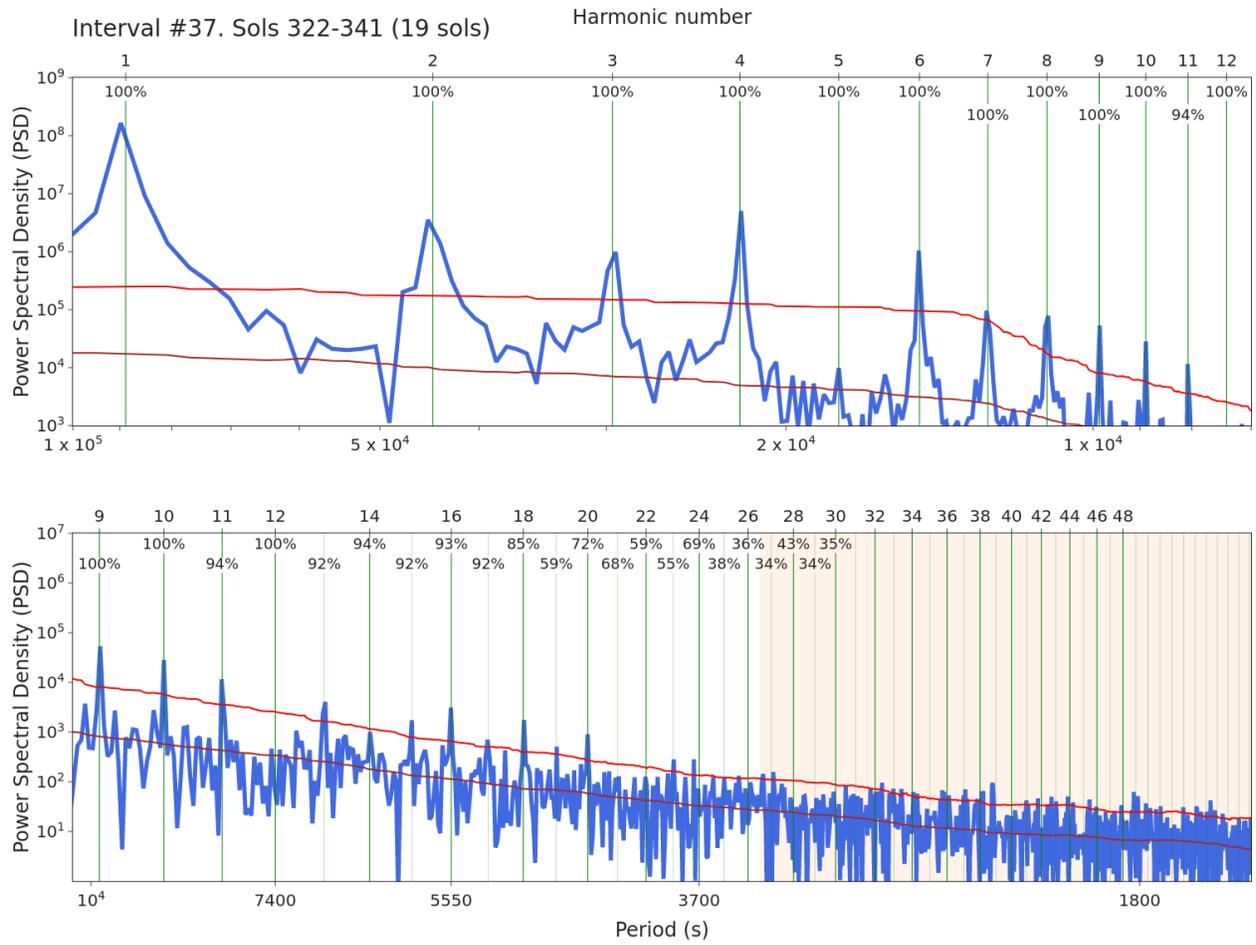

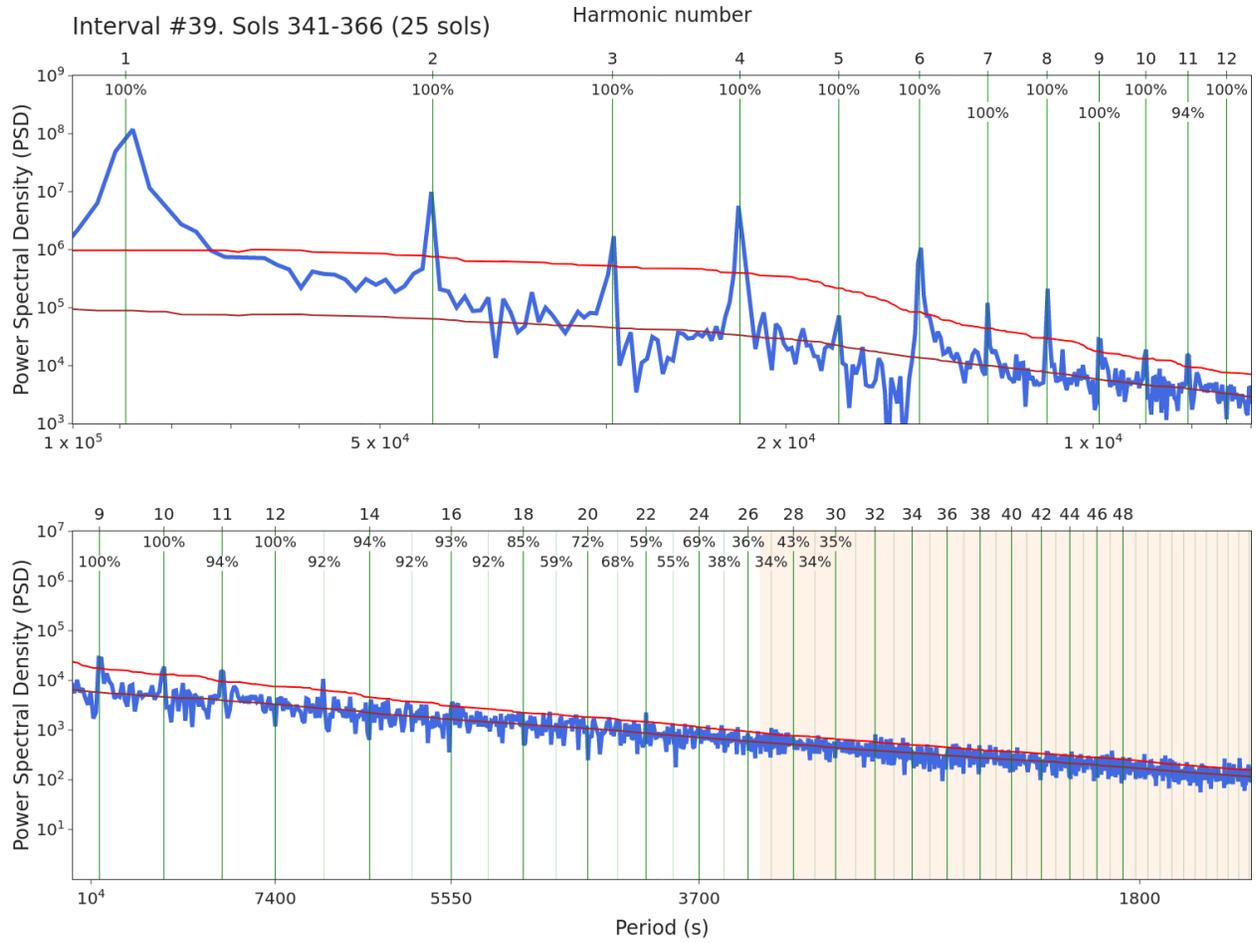

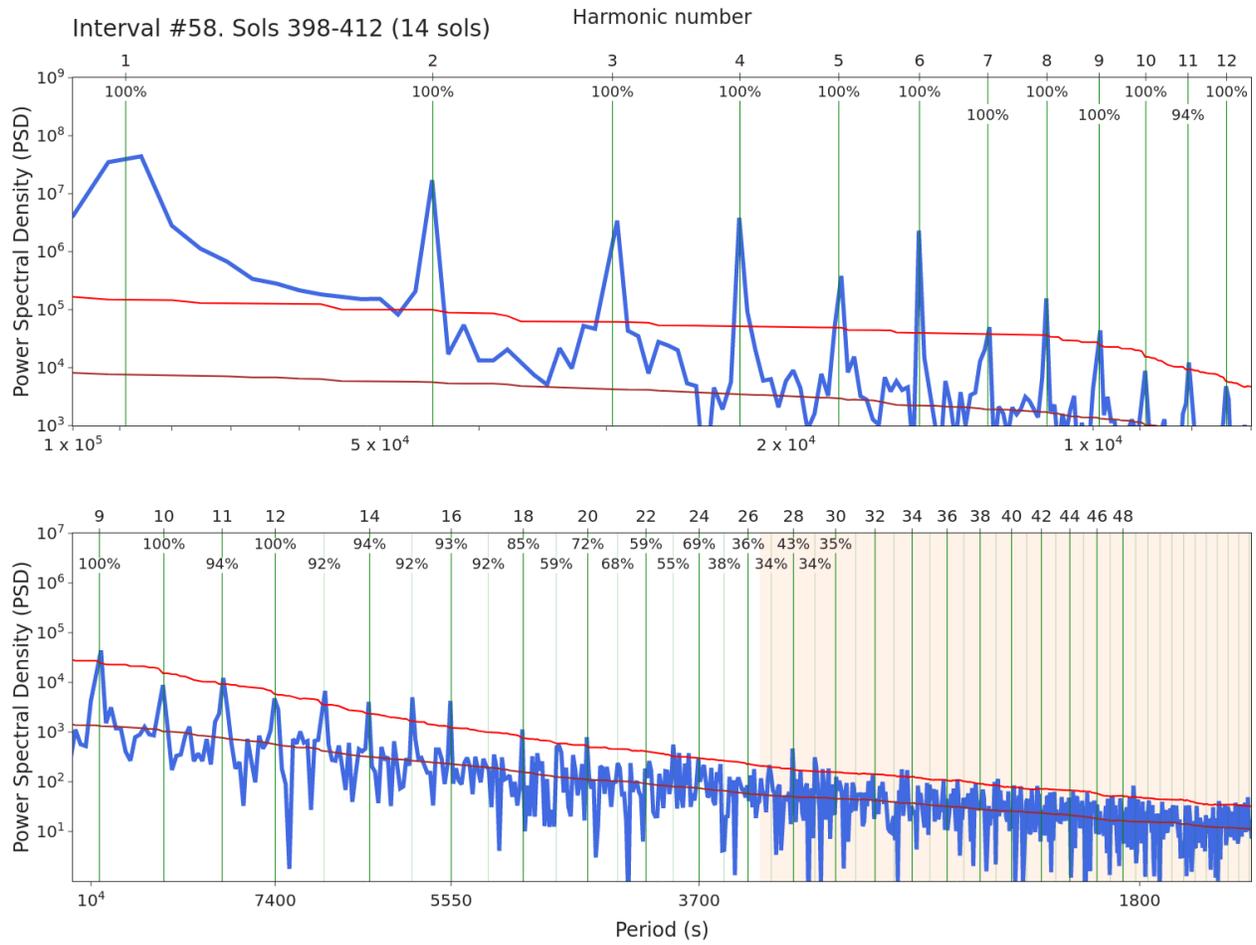

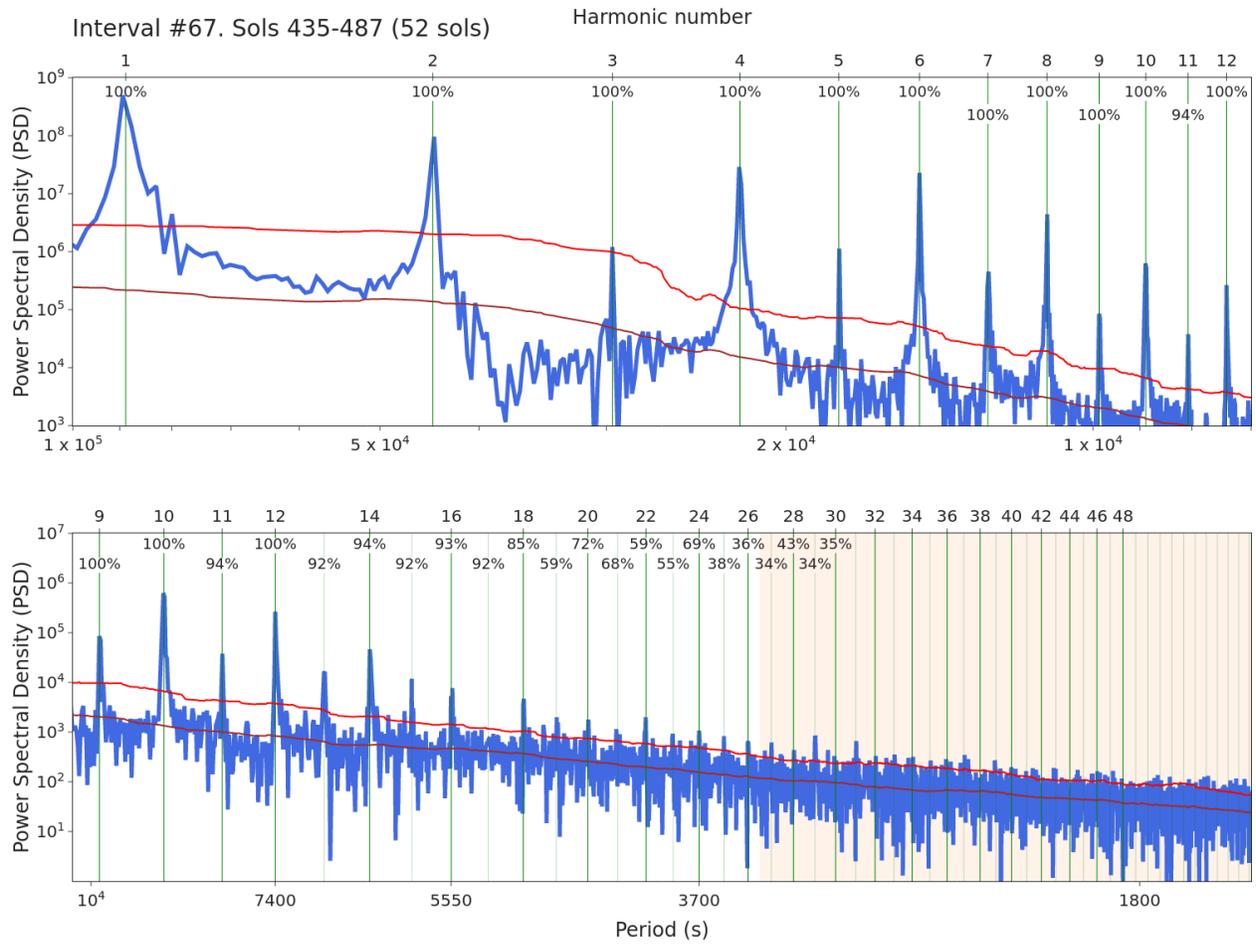

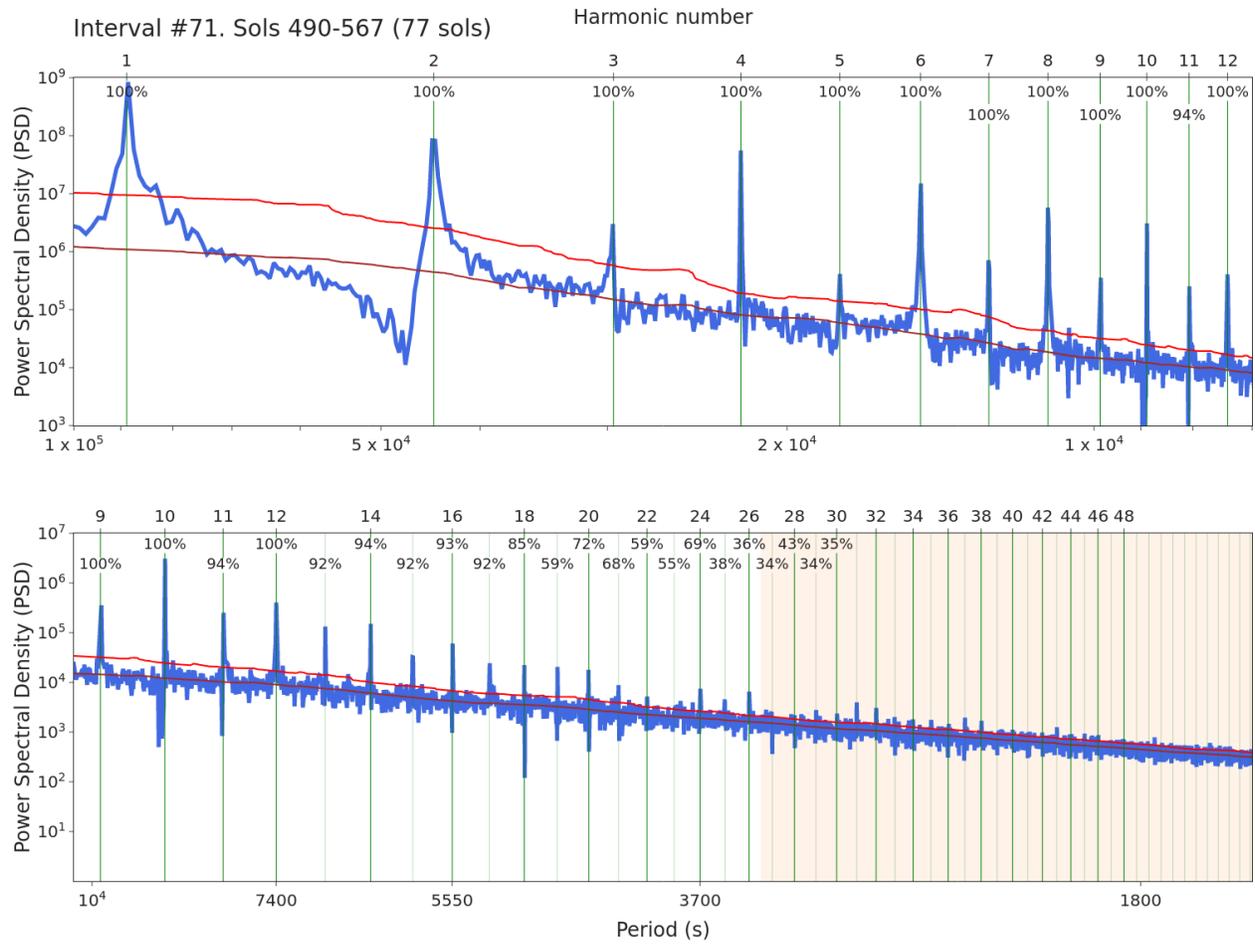

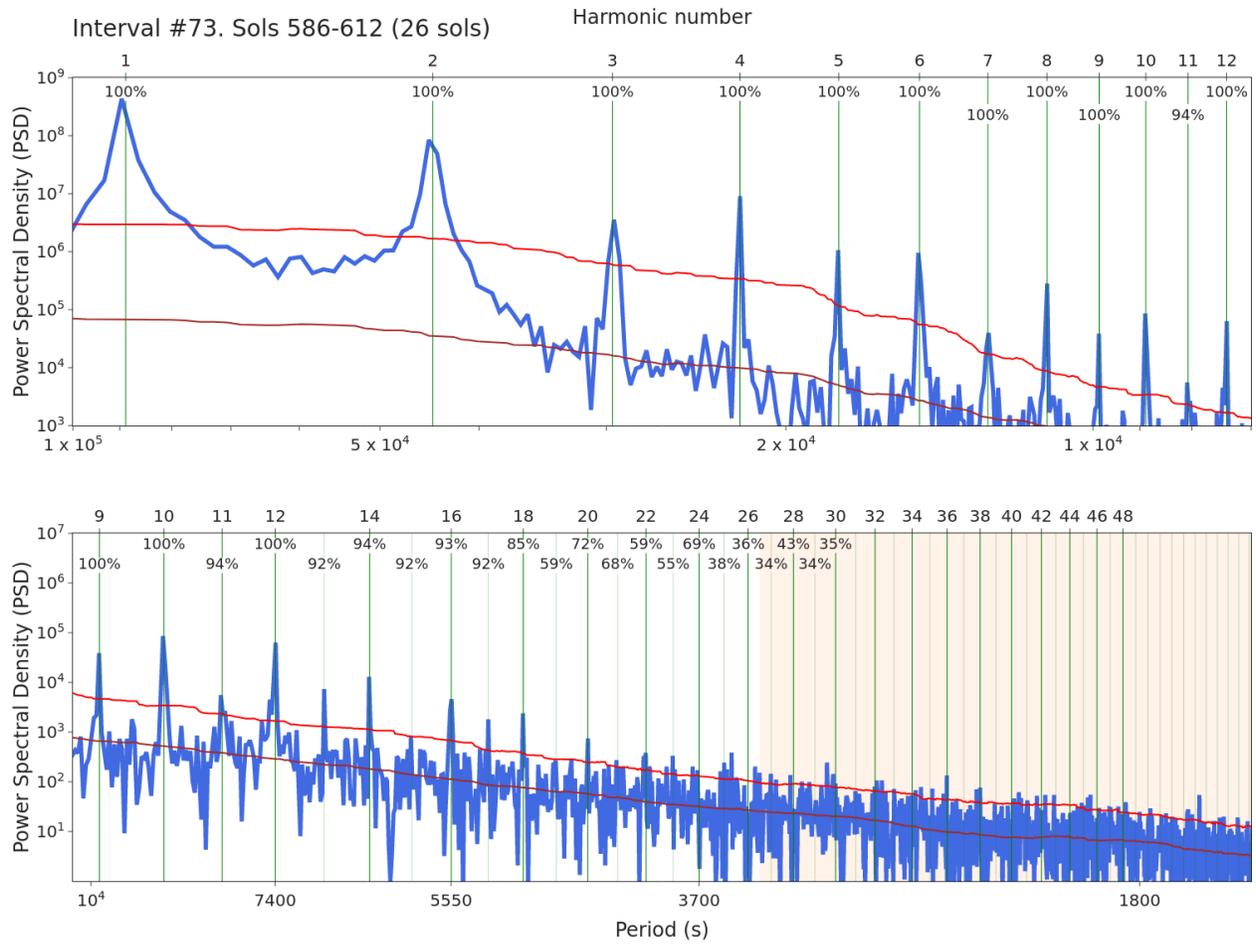

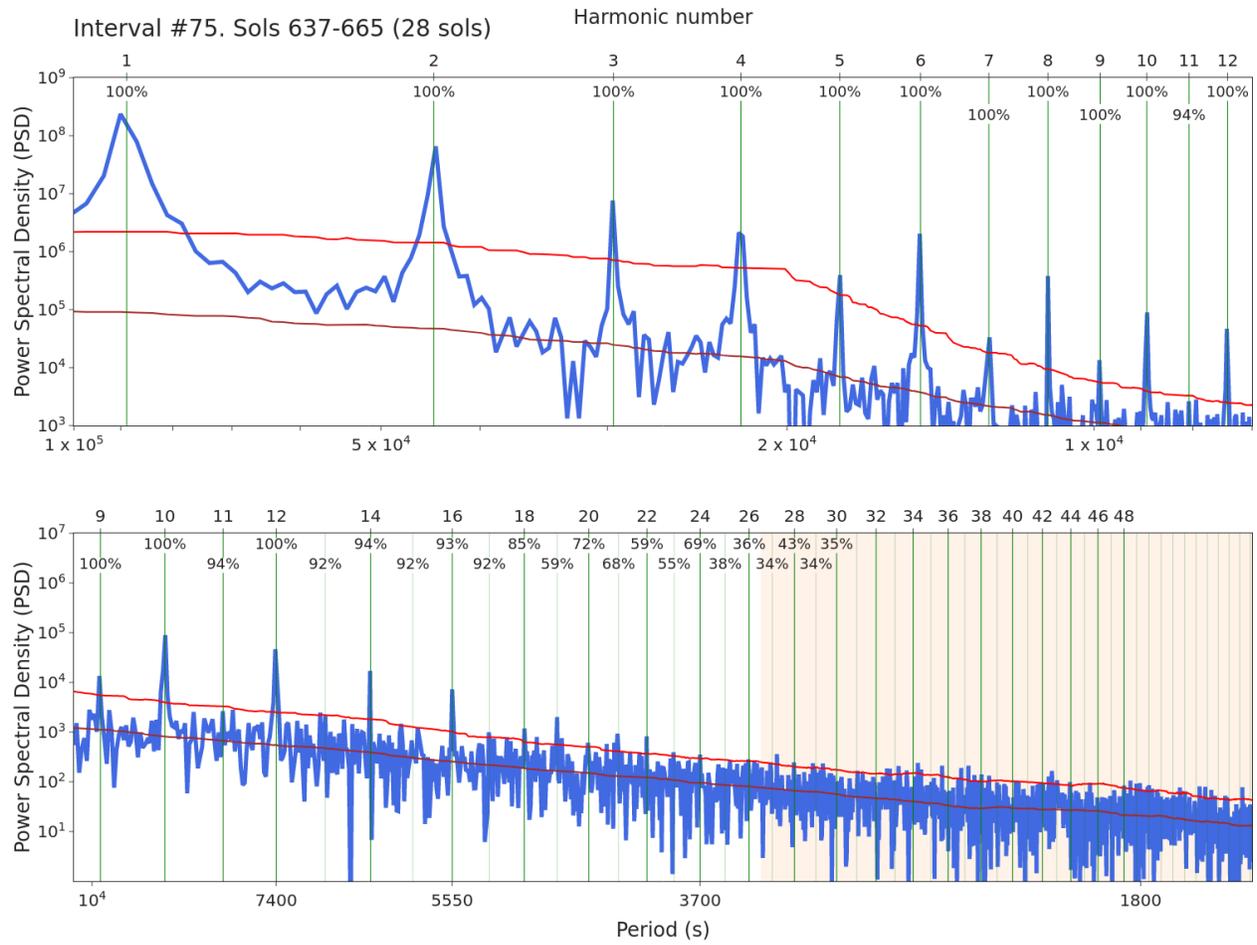

Paper accepted for publication in Geophysical Research Letters https://doi.org/10.1029/2023GL107674
This is a free version with an independent format and identical contents.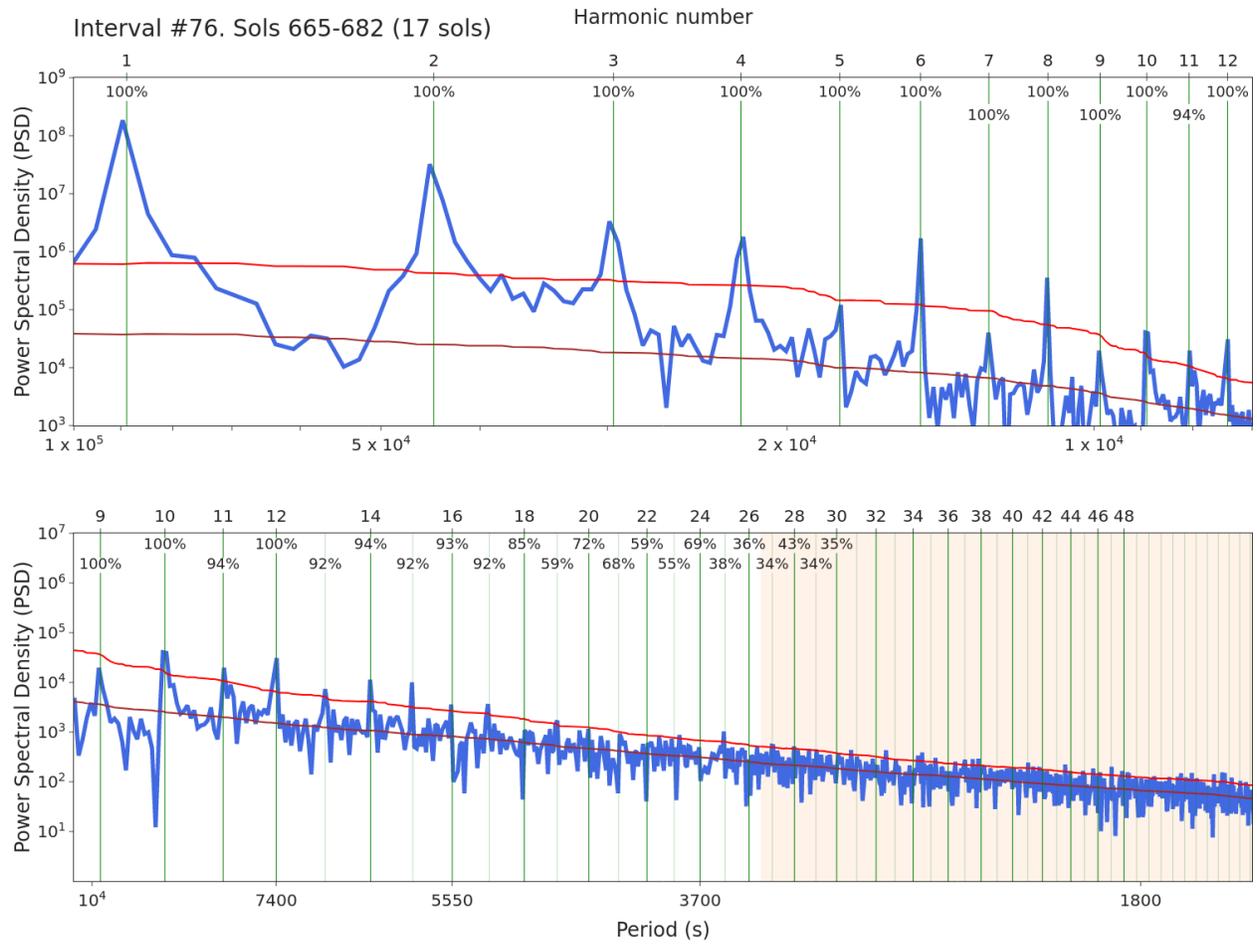

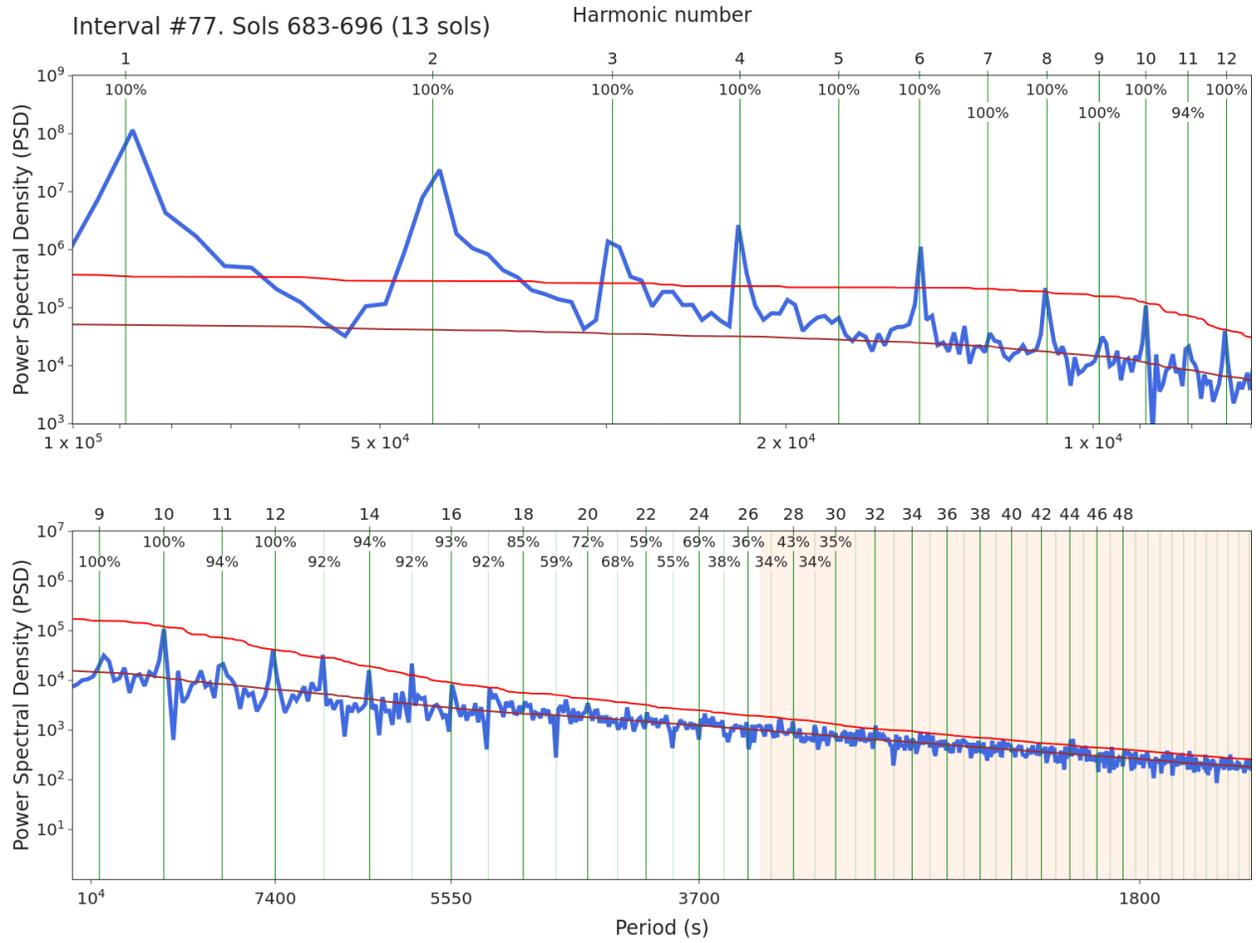

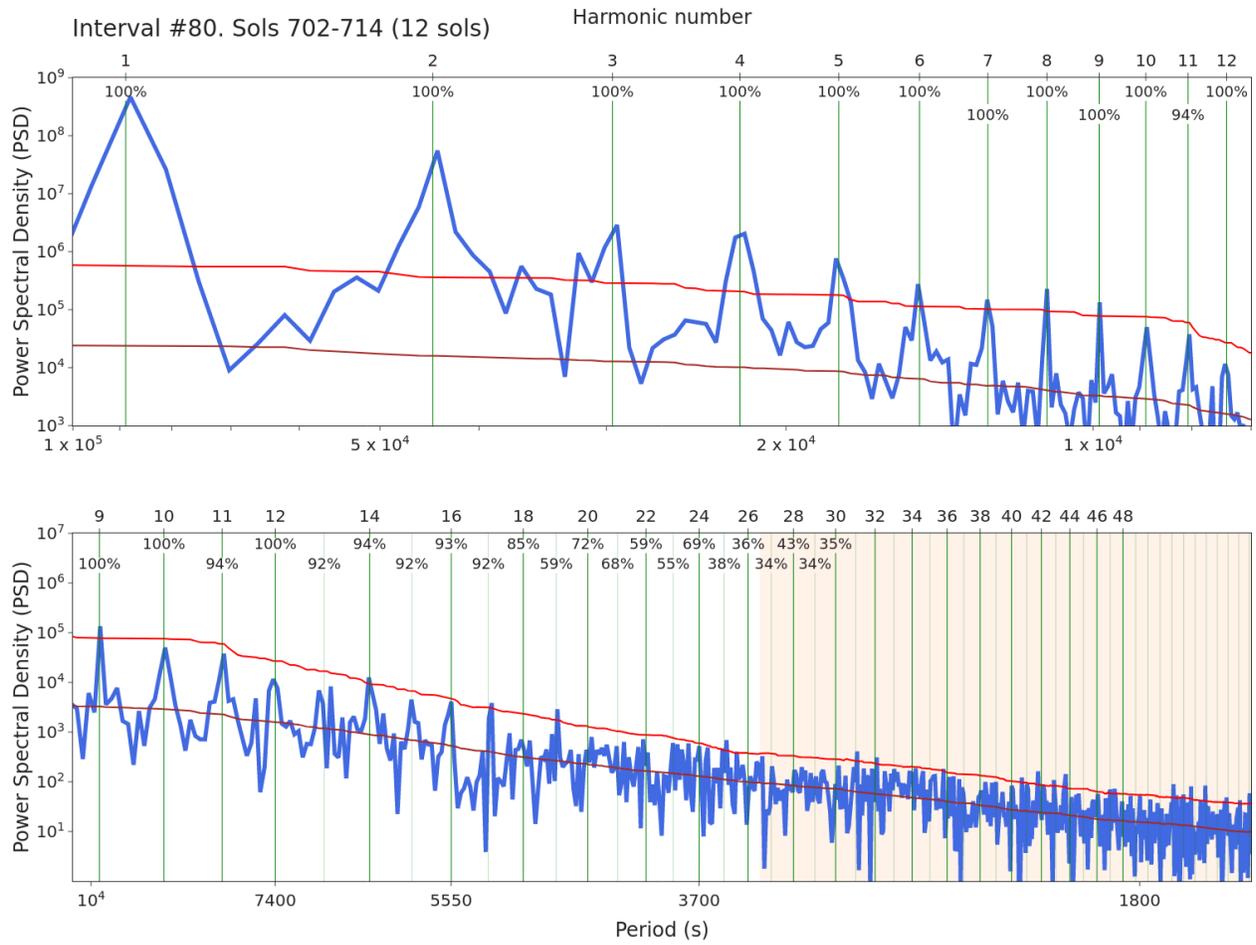

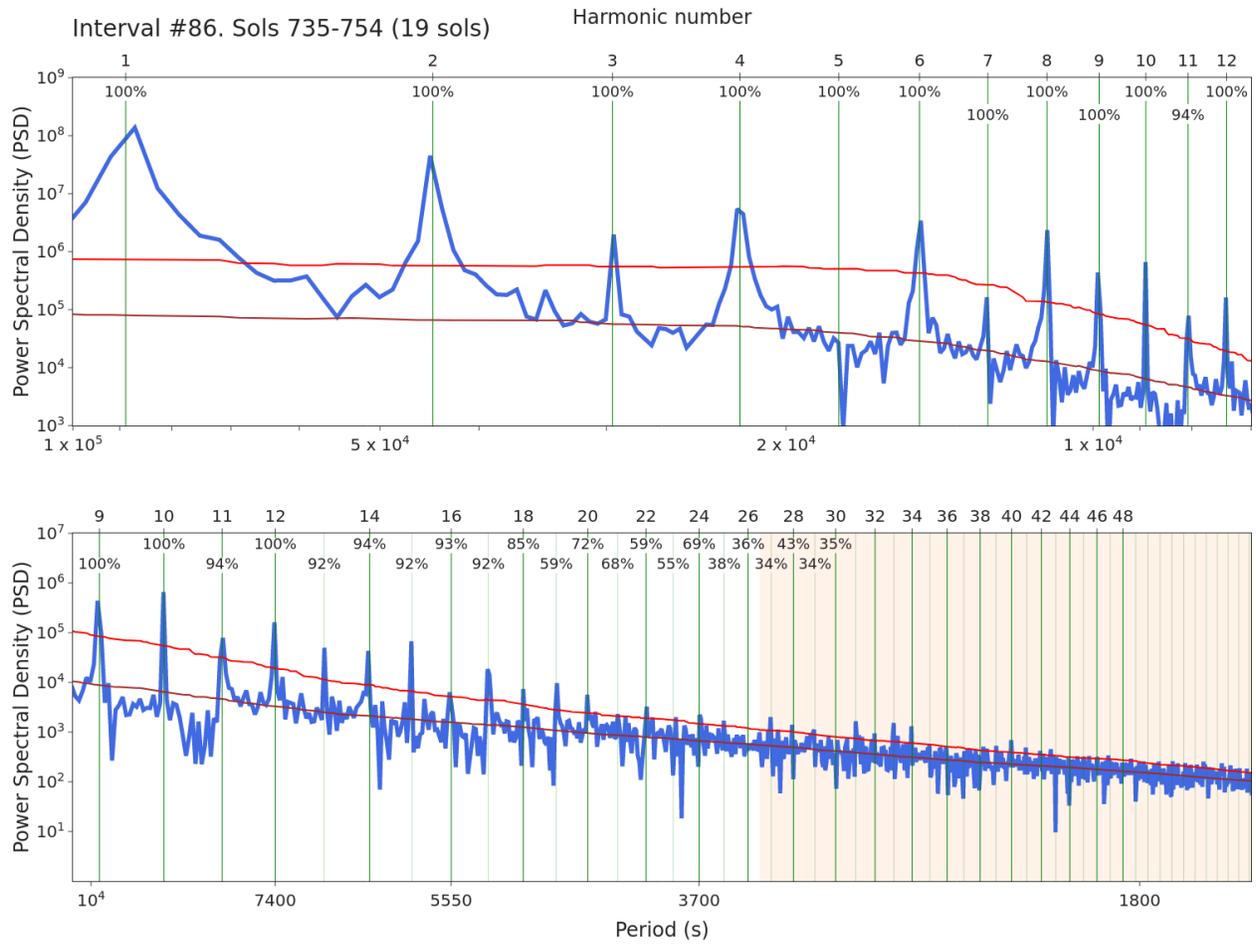

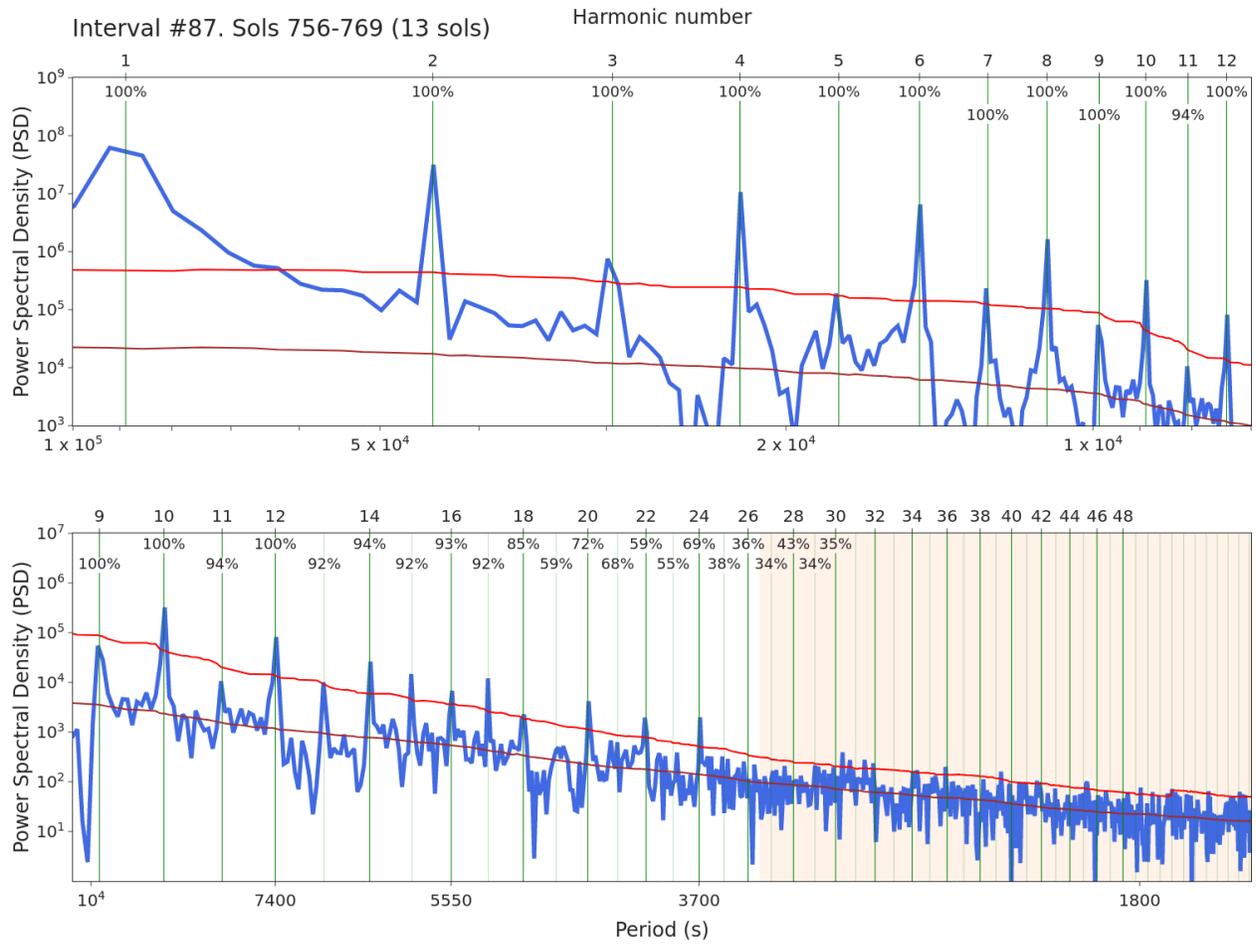

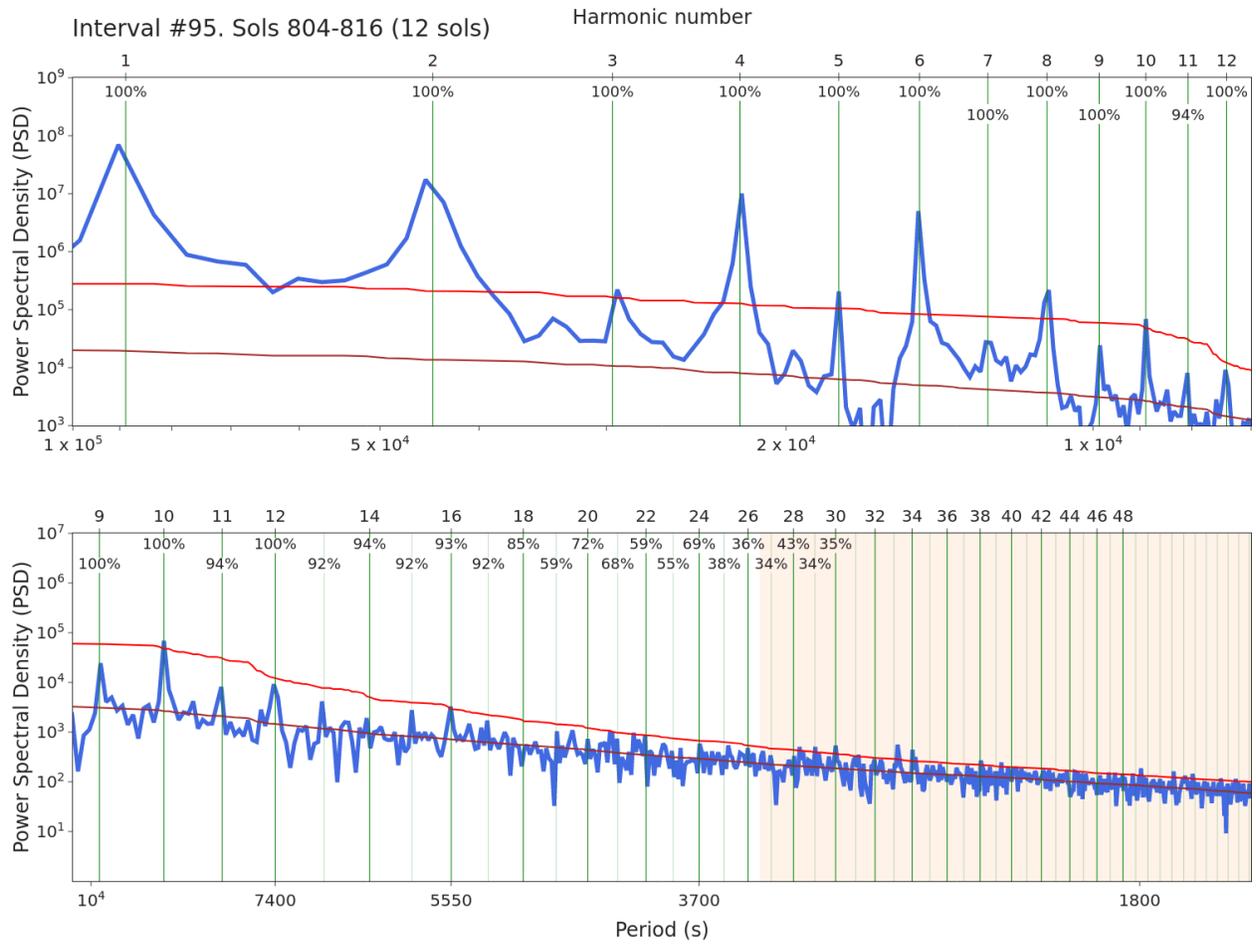

**Text S2. Computation of the estimated noise level in fig. 1 and supporting figures S2.**

In figure 1 we show the average value and an indicative estimation of the noise level. The indicative noise level corresponds to 3 sigmas in the case of figure 1 (in the case of supporting figures S2, periodograms are in general noisier, and therefore we chose the represent 2 sigmas). The calculations were done in logarithmic scale. The average and sigma are computed for each point for a window of 240 values around each point. This window is symmetric, and that removes the linear trend present in the data. In order to reduce the effects of the peaks on the estimation of the average and sigma for each point, we first compute an estimation, then remove points 2 sigmas (1 sigma in the case of the supporting figures due to the higher noise relative to the peaks) outside the average based on that first estimation, and then we use the remaining points to compute the final values of average and sigma. All the computations use numpy (Harris et al., 2020).

```python
import numpy as np

MARS_MEAN_DAY=88775.2439853306

sigma_remove=2 #1 for supporting figures
sigma_noise=3 #2 for supporting figures

P,S=np.load('Dataset S1. Periodogram for interval ID71.npy')
xx=[] #Where periods will be stored
yy_ave=[] #Where average will be stored
yy_sig=[] #Where sigma will be stored

#We will work on logaritmic scale, which is the scale of the graph
Slog=np.log10(S)

#loop for each point of the periodogram
for i in range(len(P)):
    Pi=P[i]#Period of the point
    Ni=MARS_MEAN_DAY/Pi #float number of harmonic
    if Ni<0.8 or Ni>60:continue #ignore very low and very high harmonics
    #Window for computation
    le=120
    i1=i-le
    i2=i+le
    #If the window is outside the dataset, ignore those values not available
    i1=max(0,i1)
    #
    #Points within the window
    Slogi=Slog[i1:i2]
    #Initial estimation of average and sigma
    ave0=np.mean(Slogi)
    sig0=np.std (Slogi)
    #Remove values 2 sigmas outside the initial estimation.
    Slogi2=Slogi[np.abs(Slogi-ave0)<sig0*2]
    #Then compute average and sigma
    ave=np.mean(Slogi2)
    sig=np.std (Slogi2)
    xx.append(Pi)
    yy_ave.append(ave)
    yy_sig.append(sig)
## RESULTS:
xx=np.array(xx)
yy_ave=np.array(yy_ave)
yy_sig=np.array(yy_sig)
```

**Figures S3. Analogue to Fig. 2a, for each meteorological season.**

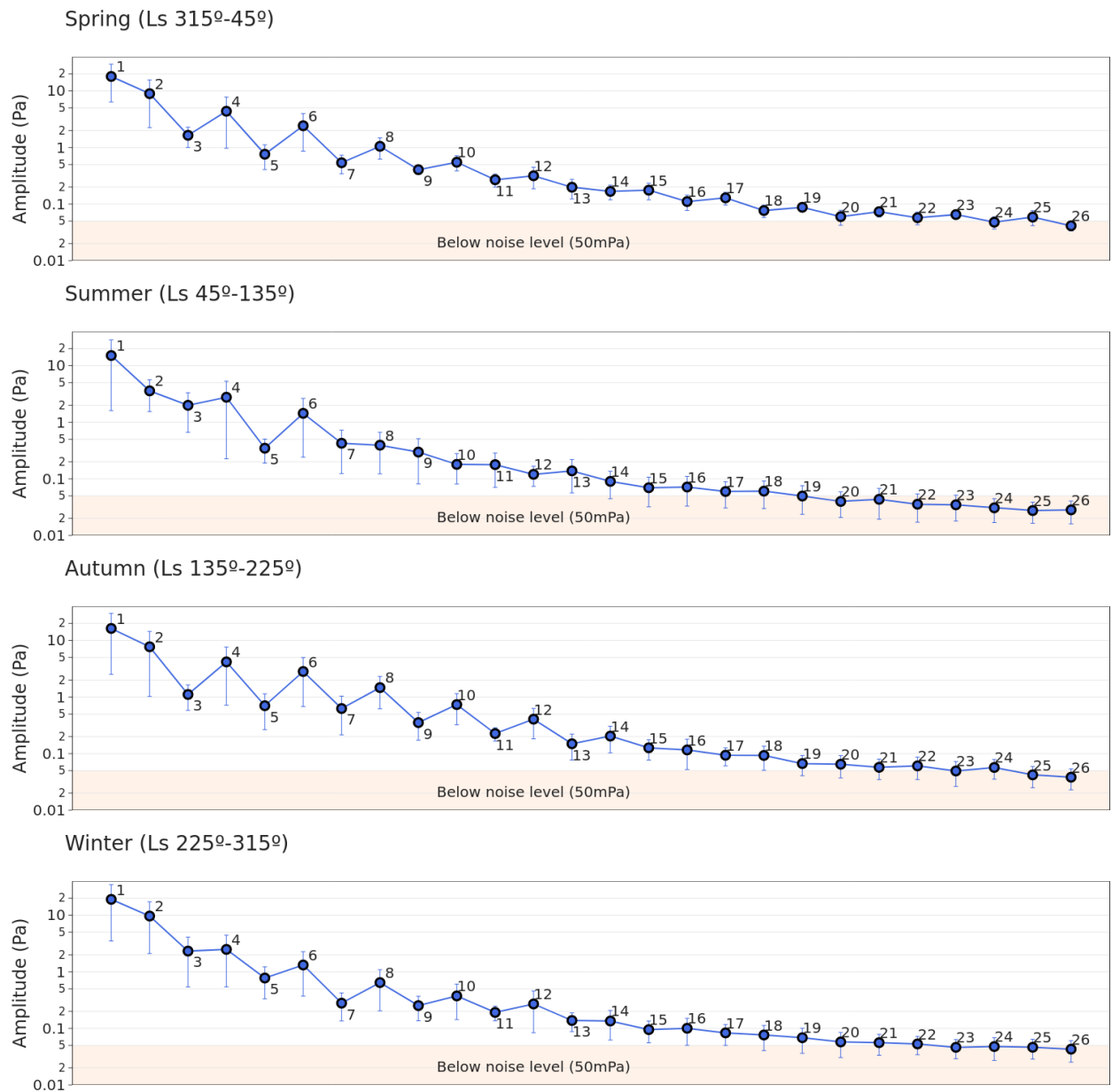

**Caption for video S3.** Equivalent to fig. 2a, for each sol. Separate sols can be analyzed using the right software, for example, the "gifview" packet in linux.

**Figure S4. Analogue to fig. 4 for harmonics 13-27.**

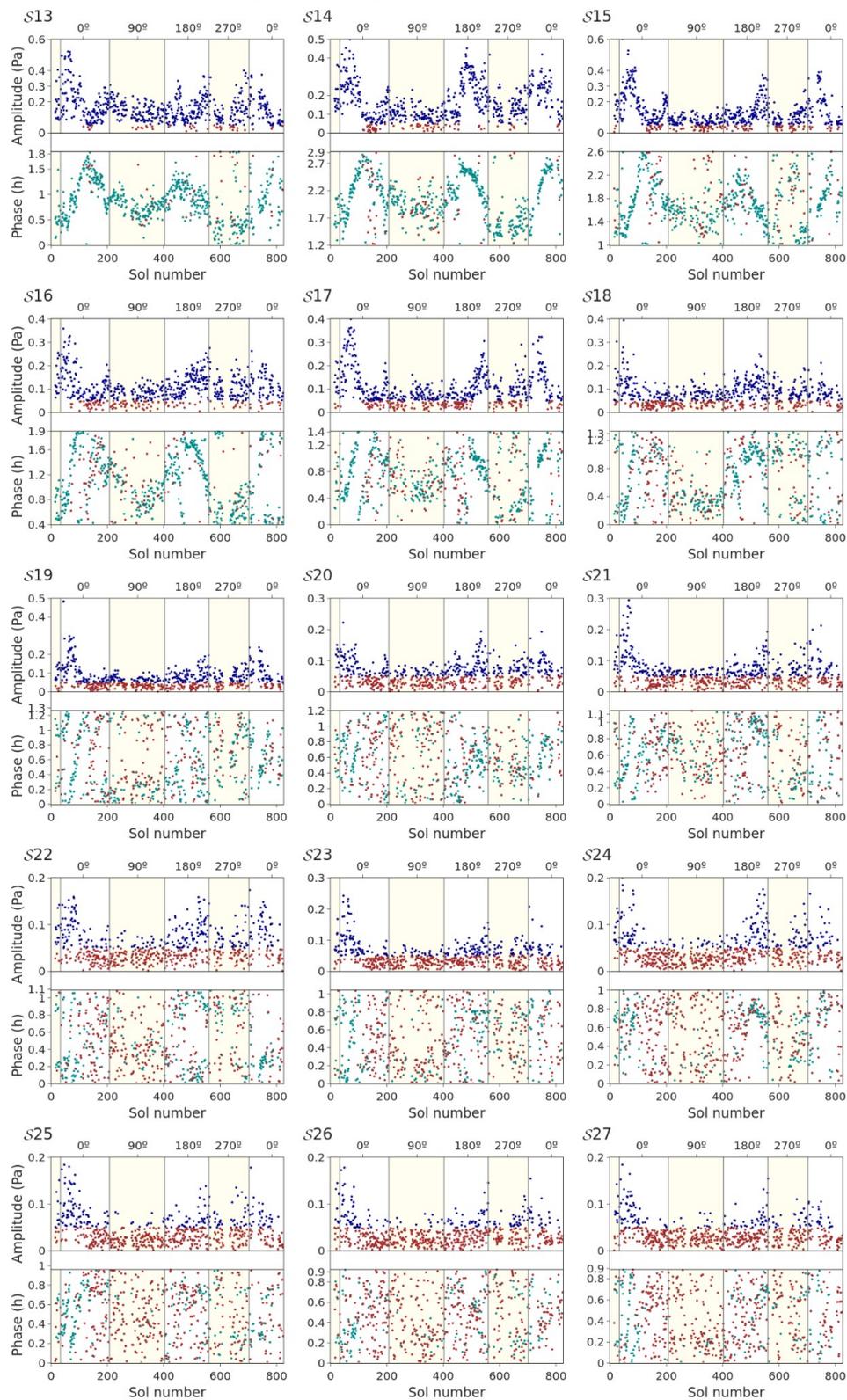

**Text S3. Details on the datasets**
All these datasets are provided in .npy format, which is an efficient way to store structured arrays provided by the python numpy library (Harris et al., 2020). It can be opened simply by executing:

> import numpy as np
> dataset=np.load('filename.npy')

Each array is codified using a numpy type, for example:

> In: dataset.dtype
> Out: dtype('float64')

Each array has also a shape:

> In: dataset.shape
> Out: (825,1440)

Note that when no data is available for a given point (usually when the sol is not available) the stored value is NaN.

**Caption to Dataset S1. Periodogram for interval ID71.**
The shape of this array is (2, 7580). The first element dataset[0] is the list of periods in seconds. The second element dataset[1] is the list of PSD values for each period (computed using numpy; Harris et al., 2020). The type of this array is float64.

This dataset is the source for fig. 1.

**Caption to Dataset S2. Ls for each sol of the Insight mission**
The shape of this array is (2, 444). The first element dataset[0] is a list of sols. The second element dataset[1] is a list of Solar Longitudes (Ls) corresponding to those sols. The type of this array is float64.

The only purpose of this dataset is to facilitate the analysis of our results by other researchers, but it can be easily computed. For example, using this library: https://github.com/eelsirhc/pyMarsTime

**Caption to Dataset S3. Tidal parameters for every sol.**
The shape of this array is (825,100). The first dimension is the number of sol (dataset[15] corresponds to sol 15, we ignore the position number 0). The second dimension is the number of harmonic (again, we ignore the position number 0, so that dataset[15][1] corresponds to H1 in sol 15). The type of this array is complex128, meaning that it contains complex numbers, which are the output of the FFT (computed using numpy; Harris et al., 2020). Storing only the first 100 harmonics is an arbitrary decision. In each case, the absolute value of the complex number corresponds to the amplitude of the harmonic in Pascals; and the phase of the complex number refers to the phase of the harmonic. They can be computed like:

```
import numpy as np
dataset=np.load('Dataset S3. Tidal Parameters for every sol.npy')

sol=129 #Sol number
Hn=1 #Harmonic number

#computation of the amplitude (Pa)
amplitude=np.abs(dataset[sol,Hn])

#computation of the phase (hour of one maximum)
angle=(-np.angle(dataset[sol,Hn]))%(2*np.pi)
phase=angle/(2*np.pi)*24./Hn
```

This dataset is the source for panels in fig. 2 and fig. 4.

**Caption to Dataset S4. Pressure derivative.**
The shape of this array is (825,1440). The first dimension is the number of sol (dataset[15] corresponds to sol 15, we ignore the position number 0). The second dimension is the Local True Solar Time, where the item number 0 represents the starting of the day, and 1439 represents the ending of the day. The type of this array is float64.

This dataset was computed using numpy (Harris et al., 2020). High frequency signals (period<3700s) were removed using a bandpass filter. Then the derivative was computed from a spline interpolation. For computational efficiency (computation time and storage) everything was downsampled to 1440 local times.

This dataset is the source for fig. 3.